\documentclass{aa}
\usepackage[varg]{txfonts}
\usepackage{graphicx}
\pdfoutput=1

\begin{document}

\title{{\bf Dust cleansing of star-forming gas I: Did radiation from bright stars affect the chemical composition of the Sun and of M\,67?}} 

\author{Bengt Gustafsson \inst{1, 2}}

\offprints{Bengt Gustafsson, \email{bengt.gustafsson@physics.uu.se}}

\institute{Department of Physics and Astronomy, Uppsala University, Sweden
  \and Nordic Institute for Theoretical Physics (NORDITA), Stockholm, Sweden} 

\date{Received ...  / Accepted ...}

\abstract {} {The possibility that the solar chemical composition, as well as the similar composition of the rich 
open cluster M\,67, have been affected by dust cleansing of the pre-solar/pre-cluster cloud,
due to the radiative forces from bright early-type stars in its neighbourhood, is explored.} {Estimates,
using semi-analytical methods and essentially based on momentum conservation, 
are made of possible dust-cleansing effects.} {Our calculations indicate that the amounts of cleansed
neutral gas are limited to a relatively thin shell surrounding the H\,II region around the early-type stars.} {It seems 
possible that the proposed mechanism acting in individual giant molecular clouds may produce significant abundance effects for masses corresponding to single stars or small groups of stars. The effects of cleansing are, however, severely constrained by the thinness of the cleansed shell of gas and by turbulence
in the cloud, why the mechanism can hardly be important in cleansing masses corresponding to rich clusters, such as the mass of the original M\,67.} 

\titlerunning{short}
\maketitle

\section{Introduction}

The finding by \citet{Melendez09} that the Sun departs from most solar twins in the solar neigbourhood in that the composition of its surface layers is comparatively
rich in volatile and poor in refractory elements, has led to considerably discussion (for a recent review see \citet{Adibekyan17}). Although the finding, a correlation
between the chemical abundances for the Sun
relative to the twins with the condensation temperatures of the elements in a gas, has been verified by several independent studies, e.g. \citet{Ramirez09}, \citet{Gonzalez10}, \citet{Adibekyan14}, \citet{Nissen15}, \citet{Spina16}, its interpretation is
still disputed. The early suggestion by \citet{Melendez09} that it could be due to the formation of terrestrial planets in the proto-planetary nebula before the latter
was dumped onto the Sun, polluting its convection zone by dust-cleansed gas, requires either an unusually long-lived nebula or a very rapid retraction of the
solar convection zone to the surface (cf \citet{Gustafsson10}). As suggested by \citet{Onehag11} this early retraction might result from the episodic accretion scenario suggested by
\citet{Baraffe10}, if applied to the Sun but not to the twins. The alternative that the effect is due to Galactic chemical evolution, and migration of the Sun from other
galactic regions than those of the twins, was suggested and partly supported by tendencies found by \citet{Adibekyan14}, \citet{Maldonado15}, \citet{Maldonado16}, and \citet{Nissen15} 
for some correlation of the abundance effects to depend on stellar ages. Still another possibility was proposed by \citet{Gustafsson10} that the pre-solar nebula was 
cleansed early on from dust by the radiation pressure from hot stars in the neighbourhood. There are indications from the isotopic composition (primarily daughters of
$^{26}$Al and $^{41}$Ca)  that the Sun was formed in a 
relatively rich cluster with at least one near-by core-collapse 
supernova affecting the solar composition (cf. \citet{Adams10} for a summary, see, however, also \citet{Fujimoto18} for an alternative scenario), and thus with some O- or B-type stars
that could have partly cleansed the solar cloud before the Solar system was formed. In order to test this hypothesis \citet{Onehag11} and \citet{Onehag14} carried out 
detailed analyses of solar-twins and solar analogues in the rich old cluster M\,67, which is known to have an overall chemical composition close to solar, and
also an age on the order of 4 Gyr. It was found that the cluster stars indeed have an abundance profile more similar to that of the Sun than most solar twins in 
the neighbourhood which then seemed to support the assumption that the proto-cluster (and proto-sun) were indeed cleansed by
radiation from hot stars.

Since the pioneering work by \citet{Schalen39} and \citet{Spitzer41} on radiatively accelerated interstellar dust grains, various effects of such grains on interstellar gas have been explored, e.g. on the dynamics of clouds in the Galaxy (\citet{Franco91}) or on the internal structure of star-forming clouds
with H\,II regions hollowing their centers (\citet{Mathews67}, \citet{Cochran77}).  
Several relatively detailed numerical models of H\,II regions with dust-driven flows have been published. 
In addition to the static models by 
\citet{Draine11}, where the dust drift in the region was explored, resulting in piled-up dust in the surrounding outer shell of the region, dynamic simulations have been made by e.g. \citet{Arthur04}, \citet{Martinez-Gonzalez14} and \citet{Kim16}. Several studies have been 
focussed on the structure and evolution of density gradients and shock fronts in these objects (\citet{Franco90}, \citet{Shu02}) and the
structure of the photo-ionization fronts (\citet{Diaz-Miller98}). \citet{Skinner15} have explored the feed-back via dust grains
of radiation from massive star clusters embedded in giant molecular clouds.

The issue studied in the present paper is whether the cleansing mechanism proposed could be efficient enough to deplete the dust, not only in the
proto-solar cloud, but even in the cloud in which M\,67, with its initially about 20,000 stars (\citet{Hurley05}), once formed. A simple model of a giant molecular cloud with 
massive stars {bf formed in its centre} is used for this study, 
in order to set  {\it upper limits} of the amount of dust-cleansed neutral gas that could be expected. Thus, our aim is not to present a realistic model of this complex physical situation but sooner
to obtain constraints on the possibilities of cleansing large gas masses by the mechanism proposed. Our study is also a simple preparatory for possible more detailed numerical simulations, provided that the present results suggest that such studies may be warranted.

Our model will represent a star-forming cloud by an initially homogeneous
sphere of dense neutral atomic or molecular gas with dust, and with a bright light source in its centre. Initially, we shall follow how the dust is affected by the radiative pressure
from the star(s) (Section 2.1). The friction force of the dust moving relative to the gas has to be considered, and this friction also moves the gas. However, we shall
see (in Section 2.2) that the complex effects of this gas dynamics are to a considerable extent compensated for when the total amount of cleansed gas is calculated. Finally, the
motion of the ionization front around the star also has to be considered -- this ionization reaches the gas cleansed, presses it outwards and may 
stop the possibilities for stars to form
easily from that gas (Section 2.3). Unlike the model studies referred to above, the present work focusses on the dust in the
neutral gas, outside the H\,II region. In Section 3 we shall draw conclusions concerning how great masses of cleansed gas that could possibly result and be available for 
forming stars.


\section{The model}
\subsection{The dust front}
In interstellar clouds, the mean free path of the molecules is orders of magnitude longer than the size of the dust grains.
In this case the equation of motion for a single spherical dust grain, illuminated by a star with luminosity $L$ can be written (see \citet{Draine11})

\begin{equation}
m\cdot \frac{d^2 x}{dt^2} = \frac{Le^{-\tau}\pi\,a^2\,\overline{q_{pr}}}{4\pi d^2 c} -  2\pi a^2 \,nkT\, F(s),  
\label{eq1}
\end{equation}
with
\begin{equation}
F(s)\approx \frac{s}{\alpha}[1+(\alpha\cdot s)^2]^{1/2} + \left(\frac{eU}{kT}\right)^2 {\rm ln}\Lambda \frac{s}{3\sqrt{\pi}/4 + s^3},
\label{eq101}
\end{equation}
\begin{equation}
s \equiv \frac{v_{\rm drift}}{\sqrt{2kT/m_{\rm H}}},  \,\,\,\alpha \equiv 3\sqrt{\pi}/8, \,\,\, \Lambda = \frac{3kT}{2a e |eU|}\left(\frac{kT}{\pi n_e}\right)^{1/2}. 
\label{eq102}
\end{equation}
\noindent  
  Here, the motion of the dust grain, with mass $m$ and radius $a$, 
in the $x-$direction is along the light rays from the star with luminosity $L$. The 
light is attenuated by extinction, and the grain is at time $t$ located at an optical depth $\tau$ and a distance $d$ from the star. 
$q_{pr}$ is a numerical factor, which is dependent on the grain composition and shape and on wavelength. For the relevant
dust and stellar light discussed here a mean $\overline{q_{pr}}$ of about 1.5 
may be adopted (\citet{Draine11}).
The speed of light is $c$ and the mass of the hydrogen atoms or molecules, whatever species assumed to dominate the gas, is $m_{H}$. 
The dust grain is retarded by a drag force, described by the second term of the right-hand side of Equation (\ref{eq1}). 
$v_{\rm drift}$ is the drift velocity of the grain relative to the gas with the number density of atoms or of molecules $n$ and the kinetic temperature
$T$. In the expression for $F(s)$ in Equation (\ref{eq101}) the last term represents the Coulomb forces, 
with $U$ being the grain potential, $e$ the electron charge and $n_e$ the
electron density. Following \citet{Draine11}, we set $|eU|/kT=2.5$.
For our neutral gas (with degrees of ionization from Table 30.1 in \citet{Draine11b}) we find the Coulomb force term to be negligible for cool 
(10 K to 100 K) gas and even rather small for warm $\sim 5000$K gas as compared with
the drag force due to the neutral gas collisions. We note that the dependence of the drag force on the drift velocity is linear for small speeds -- in typical cases with $T = 10$\,K the shift 
to a mainly quadratic dependence occurs around $v_{\rm drift}\sim 400$ m/s which is higher than typical sound speeds in the cool neutral gas. Below, we shall
call the assumption of a linear dependence of the drag on the drift velocity {\it The Small Velocity Approximation}. 

For the cases to be discussed here, the free mean path of the gas molecules is large enough, and the dust velocity most often
small enough for the first term within the parentheses
in the R.H.S of Equation (\ref{eq101}) to be dominating.
Often also the acceleration term, the first term in Equation (\ref{eq1}), may be neglected and we have:

\begin{equation}
F(s) = F_1(v_{\rm drift}) = \frac{Le^{-\tau}\,\overline{q_{pr}}}{8\pi d^2 c\,nkT},
\label{eq201}
\end{equation}

\noindent which then gives $v_{\rm drift}$ from Equations (\ref{eq101}) and (\ref{eq102}).  We note that $v_{\rm drift}$ is then independent of the grain size $a$, except for a possible weak dependence through $\overline{q_{pr}}$.

In the idealized 1D case of a fully homogeneous gas, and an initially equally homogeneous dust distribution in the gas, a stellar flux that is suddenly 
switched on may be expected to cause the dust to assemble in a narrow dust front: 
grains that are first hit by the incoming radiation are rapidly accelerated until they reach an
equilibrium with the gas friction, at drift velocities of perhaps several hundred meters per second (dependent on the illuminating star, 
its distance, the extinction of its light and the gas density). When the optical depths increase due to shielding by the dust passed, and the available momentum flux from the light source is distributed on an increasing number of swept-up dust grains, the velocities are successively retarded to typically just a few meters per second. Dust grains lagging behind the dust front are exposed to the full illumination by the star, and will immediately be accelerated and soon reach the front and there loose much of the driving momentum. An interesting aspect of this is that the bigger grains, with greater inertia, may be expected to reach further into or across the dense front before they are halted by friction. Thus, some separation of dust of different sizes may occur. Dust grains ahead of the front, with initially zero drift relative to the gas, will similarly be affected differently: the bigger grains should be passed by the front and be more slowly accelerated towards it, again due to their inertia. Thus, in this simplified model the big grains will be more dispersed across the front than the small ones. 

\begin{figure}
\resizebox{\hsize}{!}{\includegraphics{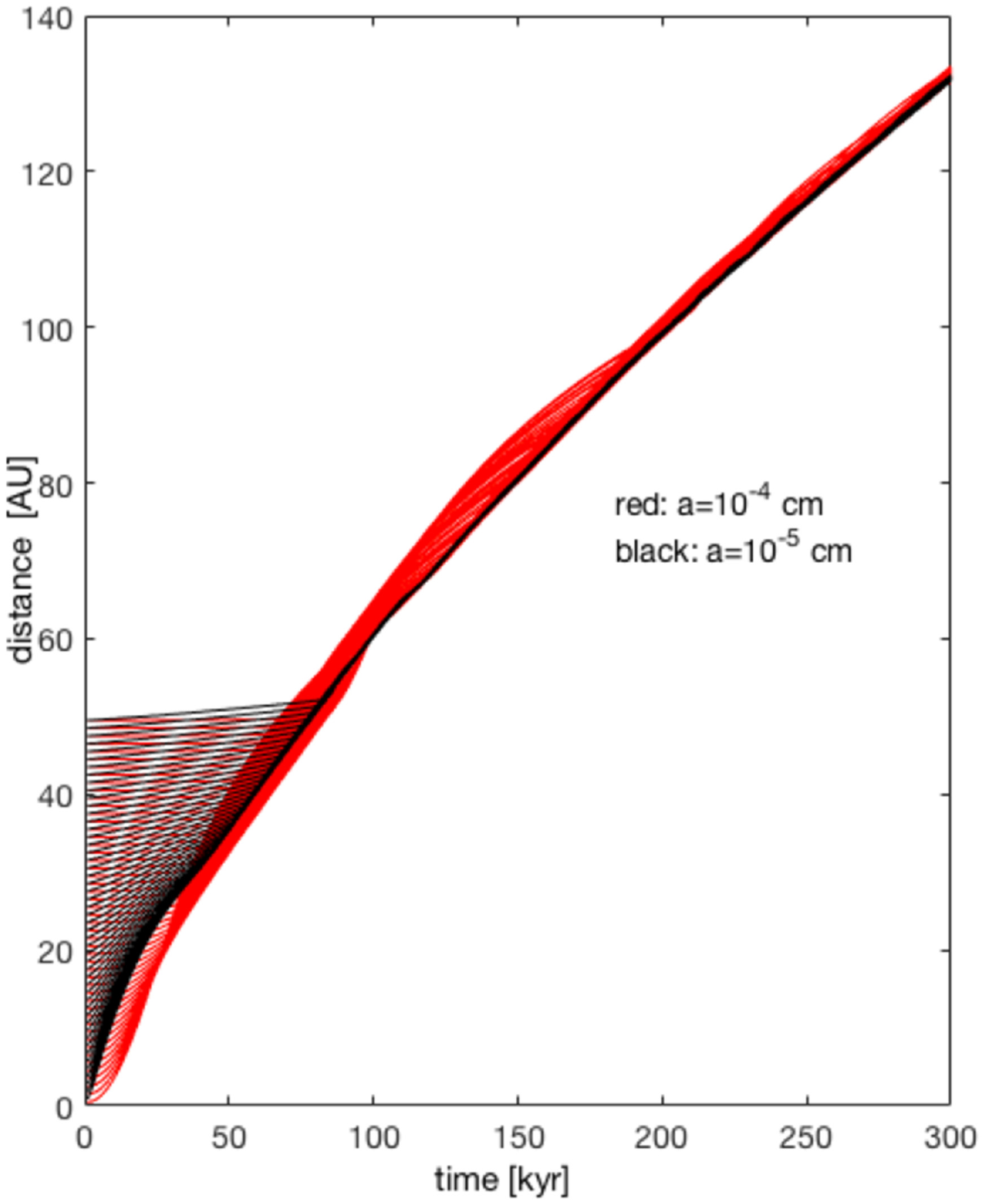}}
\caption{The motion of a stack of 100 dust sheets, initially extending across 50 AU, moving due to perpendicular radiative pressure (from below in 
 the figure) through a homogeneous gas. The motions of small ($10^{-5}$ cm) and big ($10^{-4}$ cm) grains are demonstrated by different colours.}
\label{dustfront1}
\end{figure}

\begin{figure}
\resizebox{\hsize}{!}{\includegraphics{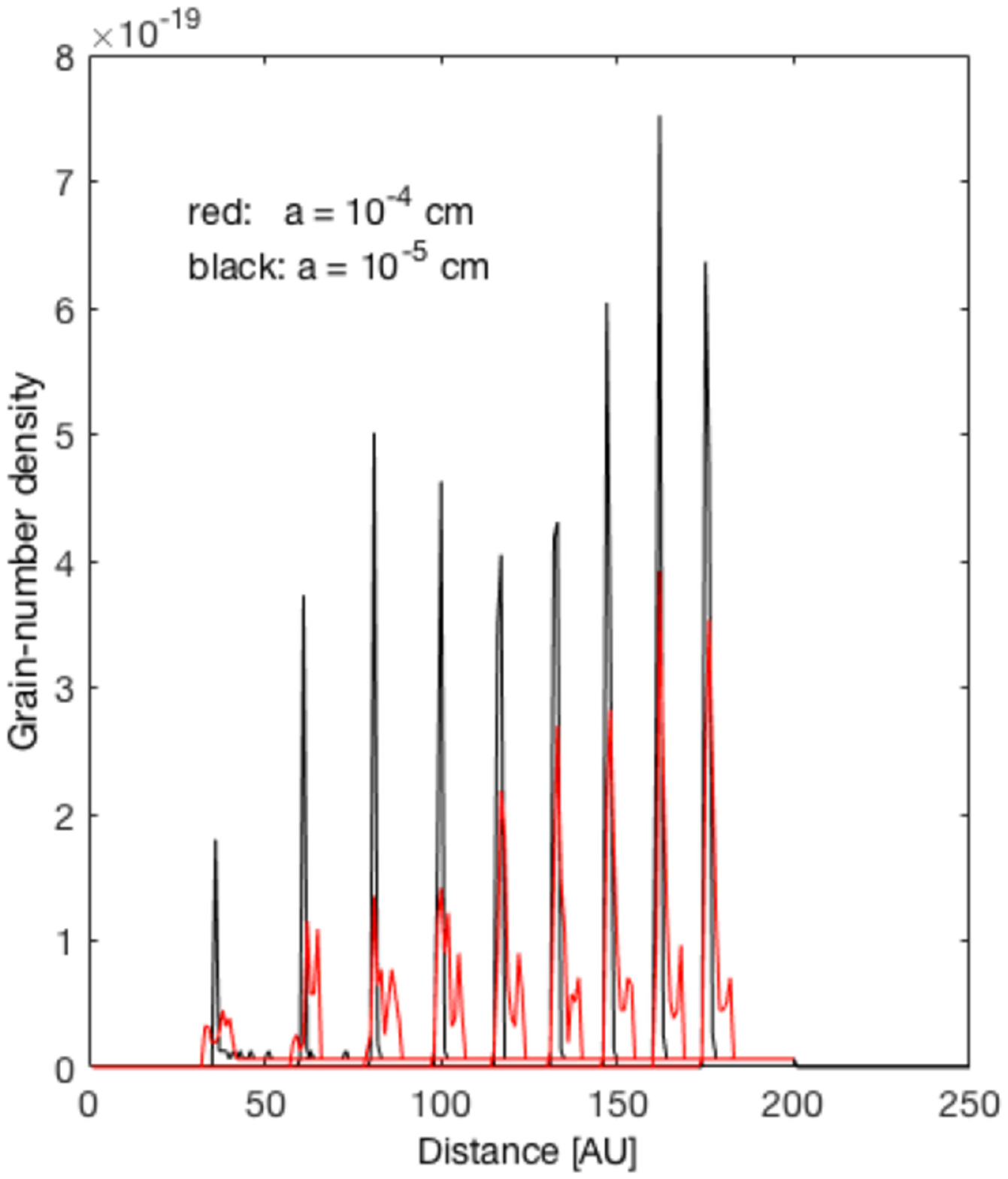}}
\caption{The dust density in the dust-front shown in Fig. \ref{dustfront1}, at 9 different consecutive times with intervals between them of 50 kyr, for two different grain sizes.}
\label{dustfront2}
\end{figure}

To explore this further, we have modeled the dust distribution by solving Equation (\ref{eq1}) for a stack of plane-parallell dust sheets, of equal masses, that are pushed along in a homogeneous and steady gas column by a light source, with a flux perpendicular to the plane of the sheets. Each dust sheet is free to move through the others and the relevant changes in the optical depths $\tau$ are successively taken into account.  To be specific, we set the dust mass density $\rho_d = 1$ g/cm$^3$ and the gas-to-dust mass ratio $\phi=100$. We switched on an illuminating star at a distance $d$ from the column. Two different sizes of grains were explicitly considered, with radii $a$ of $10^{-4}$ and $10^{-5}$ cm, respectively, with equal total mass contributions, and the combined effects {\bf were} included in the calculation of $\tau$. The density of H$_{2}$ molecules $n=10^{5}$ cm$^{-3}$ and for the stellar luminosity and distance $L/d^2 = 10^{38}$ erg/s, pc$^{-2}$. Typical results are shown in Figs \ref{dustfront1} and \ref{dustfront2}. In the first Figure we illustrate the evolution of a dust region just 50 AU thick (limited in size for clarity) and imbedded in an extensive homogeneous and static gas cloud. It is seen how the smaller (black) grains gather in a thin decelerating dust front while the bigger (red) grains start more slowly why they tend to lag behind the smaller ones during the first 50 kyr. Then they pass the front and lead the front although the small grains gradually catch up. In Figure \ref{dustfront2}, representing the evolution of the illuminated surface layers of an extended dusty gas, the distribution across the front for the two types of grains is shown for a number of times with intervals of 50 kyr. As is seen, the leading more massive grains actually show a two-peak distribution: with a first peak ahead of the small-grain dust front by at first typically 5 AU and next successively thinner, and the other and more massive one more aligned with the front of the small grains. Obviously, the front may be totally about 10 AU wide for big grains (that have diameters of $\sim 1$ micron in typical cases), while the front of the smallest grains ($<0.02 $ microns) is just a few AU. 

The picture presented above is, however, too simplified: the front may have a speed relative to the gas of only some meters per second, and the grains that happen to be in front of it are affected by more dynamical forces and motions, for instance related to turbulence. Thus, the dust front may be much less well defined. In addition to this, it is easy to see that the dust front may be affected by instabilities; a spurious reduction of the number of grains in a section on the front will admit radiation to push the dust (which may also drag the gas) in that section further ahead of the front. Similarly, an increase of the number of grains in a region will cause it to move more slowly. If these perturbations occur early on, in regions with moderate optical depths, they may have noticeable consequences for the shape of the front, like formation of finger-like structures out from it. 

\subsection{The motion of the dust front}
\begin{figure*}
  \resizebox{\hsize}{!}{\includegraphics{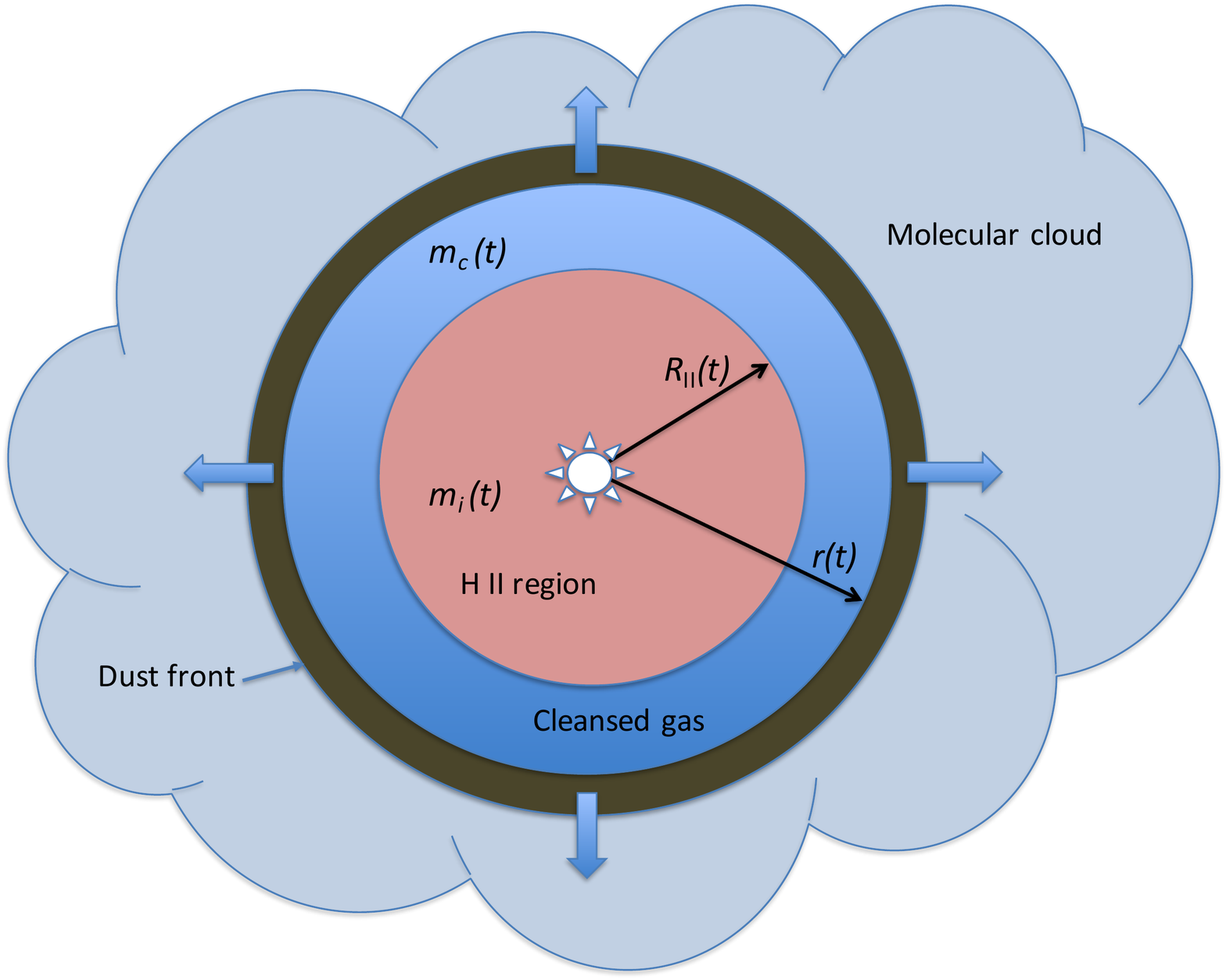}}
  \caption{A schematic picture of our model. The bright star(s) is in the middle, surrounded by the expanding H\,II region with radius $R_{{\rm II}}$. Outside the H\,II region is dust-cleansed neutral gas, surrounded by the dust front. The proportions in the figure are 
not realistic: the true thickness of the cleansed region is only a few percent of the radius of the H\,II region, and the thickness
of the dust front in turn about 1\% of that of the cleansed region.}
  \label{figxx}
\end{figure*}

We shall now estimate the velocity of the dust front through the neutral gas, and subsequently explore the resulting cleansing effects. These estimates will be made based on the assumption of spherical symmetry and the application of the momentum equation and conservation of the number of dust grains. 
We shall also make some simple approximations concerning the acceleration
phase of the dust, during its motion through optically thin layers. It  will be seen that the character of the narrow dust front discussed above will substantially simplify the discussion. A general schematic outline of our schematic model is shown in Figure \ref{figxx}.

We shall first assume the dust front to become optically thick -- departures from this {\it Optical Thickness Assumption} will be explored later. 
For the moving dust front at a distance $r$ from the star/s, the total momentum transferred by the stellar radiation to the dust per second is represented by the LHS in Equation (\ref{eq22}), below. $m_{cl}(t)$ denotes the total gas mass swept over by the dust front, and thus cleansed, at time t. 
The radiative momentum partially goes into accelerating the new dust grains caught by the front from the gas to the dust-front velocity as well as accelerating the grains already present there, resulting in the acceleration $\ddot r$.  
The total mass of this dust swept up by the front is $m_{cl}(t)/\phi$ where $\phi$ is the gas/dust mass ratio.
If the speed of the dust front is ${\rm d}r_{\rm d}/{\rm d}t = {v_{\rm d}}$ the {\it added} momentum per second to the new grains in the front (in addition to the momentum they have when swept along by the gas)
is $4 \pi\cdot r(t)^2 n(t) \cdot m_{\rm H}\mu \cdot ({v_{\rm d}}-{v_{\rm g}})/\phi$, where ${v_{\rm g}}$ is the velocity of the gas and 
$\mu =1+Y+Z$ with $Y$ and $Z$ being, respectively, the mass fractions of He and heavier elements in the gas. $n(t)$ is the local number density of gas molecules (or
atoms if the molecules are dissociated)
at the dust front. The rest  of the momentum is lost to the gas by friction, according to the last term in the following
equation:

\begin{equation}
\begin{split}
\noindent \frac{L\,\overline{q_{pr}}}{c } = \frac{m_{cl}(t)}{\phi} \cdot \ddot r + 4\pi r(t)^2 \cdot \frac{m_{\rm H}\mu}{\phi} n(t)\cdot v_{\rm d}\cdot(v_{\rm d}- v_{\rm g}) + \\
\noindent + \frac{G\cdot m_{cl}(t)}{r(t)^2 \phi}(m_{cl}(t)+m_*)+ \\
\noindent + 2 \pi kT n(t)\cdot F_1(v_d-v_g)  \int _{a_l}^{a_u} N(a,t) a^2 \, {\rm d}a.
\end{split} 
\label{eq22}
\end{equation}
\noindent $\overline{q_{pr}}$ is here dependent on the detailed radiative transfer in the dust front.
Assuming that the photons are finally absorbed and give up their total momentum in the optically thick front, we can set $\overline{q_{pr}} = 1$. 
The fourth term represents the gravity from the
gas mass on the dust shell, as well as the gravity from the stars with mass $m_*$ formed inside the dust shell at early stages before the
cleansing process started. One should note that the \textit{local} number density of H$_2$ molecules or free hydrogen atoms $n(t)$ may be quite different from the initial density, $n_0$. The temperature-dependent choice of hydrogen species will be specified later.  

$N(a) $d$a$ is the total number of dust grains 
in the dust front with radii in
the interval ($a,a+$d$a$), and $a_l$ and $a_u$ are the lower and upper limits respectively, of this dust-size distribution.
We have adopted the classical distribution $N(a,t)=N_t a^{-3.5}$ of \citet{Mathis77}, see also \citet{Casuso10}, in the range from $a=10^{-7}$ cm to $a=10^{-4}$ cm.
It is worth noting that, with this size distribution, the most important grains contributing extinction (scaling as $a^2$) are those with smaller sizes, while the mass contribution ($\propto a^3$) is more evenly distributed among the different sizes.  

We can now normalize the dust size distribution, assuming spherical grains, to the proper dust mass ratio at the time $t$
\begin{equation}
\int_{a_l}^{a_u }\! N_t a^{-3.5} \frac{4 \pi a^3}{3}\rho_d  \,{\rm d}a =  \frac{8\pi}{3}N_t\rho_d (a_u^{1/2} - a_l^{1/2}) = \frac{m_{cl}(t)}{\phi}.
\label{eq23}
\end{equation}
\noindent Since  $a_u^{1/2} >> a_l^{1/2}$  we may neglect the $a_l^{1/2}$ term in Equation (\ref{eq23}) and obtain
\begin{equation}
N_t = \frac{3}{8 \pi} \cdot \frac{m_{cl}(t)}{\rho_d \phi \sqrt{a_u}}.
\label{eq234}
\end{equation}
Correspondingly approximating the integral
in Equation (\ref{eq22}) and defining $\overline{a} \equiv \sqrt{a_la_u}$
we may now write the equation of motion for the dust front, as
\begin{equation}
\begin{split}
{\ddot r}\,+ \lambda \frac{r(t)^2n(t)}{m_{cl}(t)} v_{\rm d} \cdot ({v_{\rm d}} -{v_{\rm g}}) + \kappa T \,n(t)\ F_1({v_{\rm d}}-{v_{\rm g}}) = \\ 
-\frac{G}{r(t)^2} (m_{cl}(t)+m_*) + \frac{\phi}{c} \frac{L}{m_{cl}(t)} .
\label{eq25}
\end{split}
\end{equation}
\noindent Here
\begin{equation} 
\kappa = \frac{3}{2} \frac{k} {\rho_d\overline{a}}
\label{eq241}
\end{equation}
and 
\begin{equation} 
\lambda = 4\pi m_{\rm H}\mu  
\label{eq26}
\end{equation}
\noindent are two constants, depending on the the dust properties and the chemical composition of the gas, respectively.

\subsection{The cleansed gas mass}

It is clear from Equation (\ref{eq25}) that for a full solution we must also supply information about the gas velocity $v_{\rm g}$ 
and the gas density $n(t)$ as functions of $r$ and $t$. Another circumstance of
importance is the handling of the radiative force term in the optically thin layers of the gas where this equation must be modified. 
We shall below return to these complications but shall here first develop
a simple theory which partially circumvents these difficulties, remembering that our main ambition is to obtain an estimate of the cleansed gas mass, 
$m_{cl}(t)$, sooner than studying
the detailed dynamics of the situation. 

By direct solution with typical parameters one finds that the first two first terms of Equation (\ref{eq25}) after relatively short times,
$t_{\rm d}$, become unimportant in comparison with the third term, i.e. the dust
grains relatively rapidly acquire the drift velocity. For a linear dependence of $F(s)\sim s/\alpha$ in Equation (\ref{eq101})
one finds
\begin{equation}
t_{d} \sim \frac{\rho_d \overline{a}}{\sqrt{kTm_{\rm H}}\cdot n(t)},
\label{eq261}
\end{equation}
while for the case $F(s)\sim s^2$ 
\begin{equation}
t_d \sim \frac{\rho_d \overline{a}}{n(t)(v_d-v_g)\cdot m_{\rm H}}.
\label{eq262}
\end{equation}
Typically, this leads in both cases to $t_d \sim 10^4$ years or less and total masses of cleansed gas during this time of a few solar masses. For 
the optically thin case, similar results are obtained from Equation (\ref{eq22}) and Equation (\ref{eq2507}), below. 

The neglect of the two first terms in the equation of motion for the dust front will subsequently be called 
{\it The Drift Approximation}. In practice, with realistic values of $m_*$, we also find the gravity term in Equation (\ref{eq25}) to 
be unimportant: only for relatively compact but massive regions with $m_{cl,\odot}/(r_{pc}\cdot L_{38}^{1/2}) > 3\cdot 10^3$ the gravity term would be 
comparable to the dominating ones ($L_{38}$ being the luminosity of the central star(s) in $10^{38}$ erg/s and $m_{cl,\odot}$ being the cleansed
mass in solar masses). Also, in such cases we have found by numerical
experimentation that the gravity term does not increase the maximum value of $m_{cl}$. (There are, however, the possibilities that gravity could,
in the long run, turn the expansion of the gas cloud to a collapse, with a second phase of star formation. This will be commented on later.)
Thus, neglecting also the gravity term we get

\begin{equation}
\kappa T \,n(t)\ F_1({v_{\rm d}}-{v_{\rm g}}) \approx  \frac{\phi}{c} \frac{L}{m_{cl}(t)}.
\label{eq2501}
\end{equation}
Defining $F$ and $s$ according to Equations (\ref{eq101}) and (\ref{eq102}) we find the solution for $s$ from the resulting quadratic equation. 
Next, we obtain the gas mass passed (and thus cleansed) by the dust front in the time interval d$t$ as 
\begin{equation}
\begin{split}
{\rm d}m_{cl} = (v_d-v_g) n(t) \mu m_{\rm H} \cdot 4\pi r(t)^2 {\rm d}t = \\
= 4\sqrt{2}\pi \sqrt{kTm_{H}}\mu \cdot s \cdot n(t) r(t)^2 {\rm d}t.
\label{eq2503}
\end{split}
\end {equation}

We now combine the solution for $s$ and Equation (\ref{eq2503}) to a differential equation, from which $m_{cl}(t)$ may be obtained, if the variation of the local density
at the dust front, $n(t)$ and the distance of the dust front from the star, $r(t)$ are known:
\begin{equation}
\begin{split}
\noindent \frac{{\rm d}m_{cl}}{{\rm d}t} =\frac{4\pi}{\alpha} \mu \sqrt{kTm_{\rm H}} \times \\
\times \left[ \left\{1+\left(\frac{2\alpha^2\phi\cdot L}{c\kappa\cdot T\cdot m_{cl}(t) \cdot n(t)}\right)^2\right\}^{1/2} - 1\right]^{1/2} n(t) \cdot r(t)^2.
\end{split}
\label{eq25031}
\end{equation}

In the early phase of the expansion of the dust front the second term in the RHS of Equation (\ref{eq25031}) may dominate. In this case we find
\begin{equation}
\frac{{\rm d}m_{cl}}{{\rm d}t} = \frac{8\sqrt\pi}{3}\mu\cdot\left[\frac{\phi \rho_d \overline{a} m_{\rm H}}{c} \cdot \frac{Ln(t)} {m_{cl}(t)}\right]^{1/2}\cdot r(t)^2. 
\label{eq25032}
\end{equation}
However, as the dust front accumulates more dust, so that the stellar radiation momentum is distributed on a greater dust mass, 
the second term on the RHS of Equation (\ref{eq25031}), gets  $<<1$. Putting in realistic numbers, we find that in this case, implying a 
shift from a quadratic to a linear
dependence of the friction force on the drift velocity, occurs when   

\begin{equation}
\frac{T_{10}m_{c,\odot}n(t)_4}{L_{38}} > 100,  
\label{eq2505}
\end{equation}
where $T_{10}$ is the temperature in the dust shell in units of 10 K and $n(t)_4$ is the local
gas number density in $10^4$ cm$^{-3}$. 
This case thus applies if considerable masses ($m_{cl} >> 1$ solar mass) are to be cleansed. It obviously corresponds to The Small Velocity Approximation as discussed above. We have then
\begin{equation}
\frac{{\rm d}m_{cl}}{{\rm d}t} = \sqrt{2} \pi^{3/2} \mu \sqrt{\frac{m_{\rm H}}{kT}} \cdot \frac{\phi \rho_d \overline{a}}{c}\cdot\frac{L}{m_{cl}(t)}\cdot r(t)^2. 
\label{eq2506}
\end{equation}

We note that the problem is significantly simplified when this approximation  
can be used: the explicit dependence on density fluctuations through $n(t)$ is then cancelled. Hydrodynamics beyond the simple application of momentum conservation only 
appears indirectly through the radial distance of the dust front from the star, $r(t)$, and a possible variation of the gas temperature $T$. Also in the high-velocity case (Equation (\ref{eq25032})) we note that the dependence on density only goes as $n(t)^{1/2}$. Since $r(t)$ is very much
governed by the expansion of the H\,II region, the growth of $m_{cl}$ is thus in general not very sensitive to the details of the structure of the neutral 
gas shell.   

For the {\it optically thin gas} Equation (\ref{eq1}) applies to every individual dust grain. In the extreme completely transparent case, with $\tau=0$, the grains initially closest to the star have been pushed out to a certain distance from it by the radiative pressure and there at a given time 
define the limits of the cleansed region. With the Small Velocity Approximation and applying Equation (\ref{eq2503}) we find
\begin{equation}
\frac{{\rm d}m_{cl}}{{\rm d}t} =  \frac{3\sqrt{2\pi}}{16} \sqrt{\frac{m_{\rm H}}{kT}}\mu \frac{\overline{q_{pr}}L}{c}.
\label{eq2507}
\end{equation}
In this case, not only $v_d-v_g$ and $n(t)$ but also $r$ are cancelled, and, with a constant $T$ and $L$, the cleansed mass will grow 
linearly with time. 

 It is easily seen by inserting typical numbers into Equations (\ref{eq2506}) and (\ref{eq2507}) that the growth of the cleansed mass may be
 orders of magnitude more rapid in the optically thin case than in the thick case. This is an obvious consequence of the shadowing in the thick 
 case of radiation by the grains closest to the star, radiation that would else directly strike further grains ahead. The optical
 depth $\tau$ of the dust collected from the cleansed mass is
 \begin{equation}
 \tau = \int _{r_0} ^r {\rm d}r \int _{a_l }^{a_u} N_v(a,r) \sigma(a) {\rm d}a  = \frac{3}{16\pi} \frac{\omega}{\rho_{\rm d}\phi \overline{a}}\cdot \frac{m_{cl}}{r(t)^2},
 \label{eq1531}
 \end{equation}
where $\sigma(a)$, which we here set $=\pi a^2$, is the absorption cross section per grain and $\omega$, $1 \le \omega \le 3$, depends on the $r$ distribution of dust. 
$N_v(a,r)$ is the number of dust grains with radii in the interval $(a,a+{\rm d}a)$ per volume unit. 
The lowest value of $\omega$, chosen subsequently, corresponds to a thin dust shell with radius $r$ and the highest to the original 
homogeneous distribution. 

It is important to consider how far the dust-front proceeds through the gas before it gets optically thick. If we assume 
that the initial number density of H molecules in the gas was $n_0$ and that all dust that initially was inside radius $r(t)$ has been collected in the dust front when it reaches this radius at time t, 
we can use the condition
that $\tau < 1$ and Equation (\ref{eq1531}) to estimate an upper limit of the maximum mass cleansed during
this first phase when the dust front is optically thin. We find

\begin{equation}
m_{cl} < \frac{256 \pi}{3} \left( \frac{\rho_d \phi \overline{a}}{\omega}\right)^3 \cdot \left ( \frac{1}{m_{\rm H}\mu}\right)^2 \cdot \frac{1}{n_0^2},
\label{eq1532}
\end{equation} 
which, with characteristic numbers for the parameters leads to 
\begin{equation}
m_{cl,\odot} < 2\cdot 10^7\frac{1}{n_0^2},
\label{eq1532}
\end{equation}
where $n_0$ is in cm$^{-3}$. Since the original molecular number density in the cores of Giant Molecular Clouds is typically on the order of
$10^4$ cm$^{-3}$, we conclude that the mass cleansed in such a cloud during the optically thin phase is small. 

As a check we have modeled the dust motion in the optically thin layer of the dusty molecular cloud closest to the star by subdividing the 
region into ten different sub-shells, each with thickness corresponding to 0.2 in optical depth. For these sub-shells, we have applied Equation (\ref{eq25}), properly 
modified with a factor of $exp(-\tau)$ multiplied to $L$. In solving the equation for each sub-shell individually, we found them to pile up and merge in a dust front close to a common geometrical depth point P, and from that point and onwards we integrated Equation (\ref{eq25}) for the full front. In calculating the swept gas mass we added the different sub-layers for depths outside P. As expected, the effects of this
more detailed treatment of the radiative transfer is minor for realistic shells.

It is easy to see that the optical depth of the dust shell increases linearly with its radius: the effect
of the increase of the number of swept-up grains, varying as $r^3$, combines with that of the  dispersion of the dust 
which varies as $r^{-2}$. This ends when the border of the cloud is reached. After that the optical depth decreases in proportion to $r(t)^{-2}$. 
However, except for quite small clouds, this occurs at such late stages in the evolution of the dust shell, when the speed of the shell is so slow
and the star(s) in the centre have declined so much in brightness, that it does not affect the resulting cleansed gas mass much.

Ending the discussion on the motion of the dust front we conclude that the front may be assumed to be optically thick and the
Drift Approximation may be applied (i.e. Equation (\ref{eq25031}) may be used), while the Small Velocity Approximation with a linear 
dependence of the friction force on the drift velocity is not valid in early phases (small $r$) and strong luminosities (c.f. Equation 
(\ref{eq2505})). It remains to allow for the effects of the functions $r(t)$ and $n(t)$ on the motion of the dust front. This will need 
considerations concerning the dynamics of the H\,II region, and some discussion of the pressure equilibrium of the neutral gas.     

\subsection{The ionization front}

The evolution of the region close to the bright illuminating star (or stars) is of key significance for the dust cleansing of the surrounding cool gas.  
The expanding H\,II region may well
rapidly penetrate into the molecular cloud, heat it and thus erode the possibilities for star formation. Also, this expansion,  due to the
great pressure in the hot ionized gas ($T \sim 10,000$ K) as well as radiation pressure (\citet{Krumholz09}, \citet{Akimkin15}, \citet{Akimkin17}), pushes the cool gas with its dust front to a rapidly increasing distance from the star. The gas just outside the HII region will be compressed by the out-going shock wave to high densities 
(typically $n \sim 10^{4-5}$ cm$^{-3}$ just preceding the ionization front (\citet{Draine11}, \citet{Kim16}). We shall here apply a simple approach to
modelling the HII region in order to obtain a lower limit of $r(t)$ in Equation (\ref{eq25031}). 

A bright and hot star that forms in a hydrogen cloud with initial mass density $\rho_g$ very rapidly ionizes a region around it, generating an initial Str{\"o}mgren sphere (\citet{Stromgren39}) with a radius
\begin{equation}
R_{\rm Str} = \left( \frac{3\,L\,f\,m_{\rm H\,I}^2}{4 \pi \,\alpha_r \, X\rho_g^2}\right) ^{1/3} .
\label{eq11}
\end{equation}
\noindent Here, $\alpha_r=2.6\cdot 10^{-13}$ cm$^3$s$^{-1}$ denotes the Case B recombination coefficient for hydrogen in the on-the-spot approximation (cf. e.g. \citet{Walch12}). $X$ is the mass fraction of hydrogen in the gas, while $f$ is the fraction of the stellar flux that is directly able to photoionize hydrogen from
the ground state (i.e. with a wavelength $< 91.2$ nm). $f$ is typically $10-20\%$ for O-type stars. Next, the great overpressure in this sphere, due to its high temperature of $\sim10^4$ K with respect to its surrounding, will cause it to expand (D-type expansion). \citet{Bisbas15} have found that the time-development of the radius of the region, $R_{{\rm H\,II}}(t) $ may be approximated by a combination of two functions, 
$\mathcal{R}_{\rm I}(t)$ and $\mathcal{R}_{\rm II}(t)$, according to 
\begin{equation}
R_{{\rm H\,II}}(t) = \mathcal{R}_{\rm II}(t)+(1-0.733\cdot e^{-t/{\rm Myr}})\cdot(\mathcal{R}_{\rm I}(t)-\mathcal{R}_{\rm II}(t)),
\label{eq12}
\end{equation}
where $\mathcal{R}_{\rm I}$ and $\mathcal{R}_{\rm II}$ are solutions to two ordinary differential equations (see \citet{Bisbas15}, their Equations (8) and (11), respectively).
The authors compared this approximation with more detailed numerical simulations and found it to agree excellently with the detailed results obtained
for $t \ge 0.05$ Myr. We have solved those ordinary differential equations adapted to our model parameters and consider the cleansing process to stop if/when the HII front
reaches the dust front.  

The effects on the dust cleansing in the molecular cloud are  dependent on the early evolution of the bright star and its H\,II region, i.e. on the phase when the radiation, but not the ionization front, has penetrated into the neutral gas.  The early evolution of massive stars and their surrounding ionized gas is much
dependent on their accretion history. This has been investigated recently in a number of studies, for references see \citet{Haemmerle16}. These authors followed accretion histories of up to 10$^5$ years, where in some cases the model did not approach the ZAMS until late, and found that convective PMS objects may delay the growth of the H\,II region by up to 10$^4$ years. We have schematically explored the consequences of this by assuming a linear growth of the 
Str{\"o}mgren radius from an initial value corresponding to $L/10$ to $L$ within the time $t_{br}$, which we have varied from 0 to $10^5$ yr. 

The assumed over-all arrangement of the star, its expanding H\,II region into an H\,I region, and the H$_2$ cloud where cleansing is expected to occur is illustrated in Fig. \ref{figxx}. In our calculations, we have alternatively assumed all neutral gas to be molecular, or all to be atomic, with corresponding temperature corrections, see
below. In our calculation of the cleansed mass available for new stars
we have subtracted this ionized gas. From \citet{Bisbas15}, their Equation (7), we obtain the following mass estimate for the ionized gas $m_i(t)$:
\begin{equation}
 m_i(t)= \frac{4\pi}{3} n_0m_{\rm H}\mu R_{\rm Str}^{3/2} R_{\rm H\,II}^{3/2}, 
 \label{eq152}
 \end{equation}
 and thus get the reduced cleansed neutral mass $m_c(t)$ available for star formation:
 \begin{equation}
 m_c(t) = m_{cl}(t)-m_i(t).
 \label{eq153}
 \end{equation}

The composition and the fate of the dust in the H\,II region is an important issue: how much of the dust 
is destroyed in the hot region, how much is piled up at the border and then gradually pushed into the neutral gas,
and how much is shot far outside it
and maybe even outside the molecular cloud? 
Work has been made on related situations (see, e.g. \citet{Krumholz09}, \citet{Draine11}, \citet{Martinez-Gonzalez14}, \citet{Kim16}) but not explicitly on the effects on the dust density in the gas outside the H\,II region. Since we are here primarily interested in estimating the maximum cleansing effect, 
we assume all the dust in the cloud
to gradually be collected in the dust front. 
 

\subsection{The gas density and the radius of the dust front}

Obviously the gas plays an important role in regulating the cleansing effects, primarily through its velocity in the friction term in the equation of motion for the dust
(Equation (\ref{eq25})) but also more indirectly in affecting the density distribution and transporting mass and momentum through gas flows, sound waves and shocks. 
Since our interest is focussed on determining an upper limit on the cleansed mass and the Drift Approximation may be used, 
we do not need to determine the gas velocity directly; what matters is sooner the local gas density at the dust front, $n(t)$. In order to
estimate that we have applied the quasi-hydrostatic-equilibrium condition in the gas shell, following \citet{Rahner17} as suggested
by the observations by \citet{Pellegrini07}. Thus,
the gradient of the internal pressure in the gas, $P$, is assumed to be in equilibrium with the gradient of the deposited radiative momentum per surface area and  time unit. For any given time $t$, we may then obtain the pressure structure by solving 
\begin{equation}
\frac{{\rm d}P(r)}{{\rm d}r} = \frac{L}{4\pi c r^2} \cdot e^{-\tau} \frac{{\rm d}\tau}{{\rm d}r}, 
\label{eq2501}
\end{equation}
with the gradient of $\tau$ from Equation (\ref{eq1531}):
\begin{equation}
\frac{{\rm d}\tau_d}{{\rm d}r} = \frac{3}{8\pi}\frac{\omega}{\rho_d \phi \overline{a}}\cdot \frac{m_{cl}(t)}{r^3}.
\label{eq2502}
\end{equation} 
We solve equations (\ref{eq2501}) and (\ref{eq2502}) together with the proper differential equation (Equation (\ref{eq25031}) or
(\ref{eq2507})) and with the boundary conditions $\tau(r_0) = 0$, $P(r_0) = P_0$ and the initial condition $m_{cl}(0)=0$. We combine the solution $P(r)$  with
the equation of state to get $n(t) = P(t)/(kT)$ (for values of the temperature T, see Sec. 2.6 below). We find an 
almost linear increase of log\,$n(t)$ with log$\, \tau$, up to a constant value at $\tau \sim 1$. (This is found in practice to be 
independent on the values of $r_0$, here set to $R_{\rm Str}$, and of $P_0$.) This value, $n(\tau=1)$ is taken as representative for $n(t)$, the density in
the dust shell. 

We also need to estimate the radius of the dust shell. This is done by noting that the gas pressure should dominate in the warm ionized gas
(\citet{Lopez14}) and assuming that the gas pressures on each side of the ionization frontier are close to
identical. Thus, $n_i T_i = n_nT_n$, where $n_i$ and $T_i$ are the number density of particles (nuclei and electrons) and temperature, 
respectively, in the ionized zone and $n_n$ and $T_n$ the corresponding quantities in the neutral zone, extending from the ionization front
to the inner border of the dust front. \citet{Kim16} have shown that the piling up of ionized gas at the ionization front due to the radiative 
pressure may increase the thermal pressure at the border by as much as a factor of two or more for high values of $L$ and $n_0$,
while the gas gets highly depleted inside the H\,II region. We find,
however that such an increase of $n_iT_i$ in our boundary condition only leads to small effects on the calculated cleansed mass. We obtain
\begin{equation}
\frac{m_i(t)\cdot T_i}{\mu_i\cdot \frac{4\pi}{3} R_{\rm H\,II}(t)^3} = \frac{m_{c}(t)\cdot T_n}{\mu_n\cdot \frac{4\pi}{3}(r(t)^3-R_{\rm H\, II}(t)^3)}
\label{25031}
\end{equation} 
where $\mu_i$ and $\mu_n$ are mean molecular weights in the ionized and neutral regions, respectively.
From this we find
\begin{equation}
r (t) = R_{\rm H\,II}(t) \cdot \left[1 + 0.65\cdot \frac{m_{c}(t)}{m_i(t)}\frac{T_n}{T_i}\right]^{1/3},
\label{eq25039}
\end{equation}
and may thus estimate $r(t)$ from $R_{\rm H\,II}(t)$ and $m_i(t)$ as obtained from Equations (\ref{eq12}) and (\ref{eq152}). 

\subsection{The model}
 
In the model calculations the equation of motion for the dust front as transformed to the equation for the growth of the cleansed mass, 
alternatively Equation (\ref{eq25031}) and Equation (\ref{eq2507}) depending on the optical depth, was solved together with the equation for the ionization front, Equation (\ref{eq12}) and its two underlying
differential equations, supplied with the equations for estimating the density, optical depth and radius of the dust front, Equations (\ref{eq2501}, \ref{eq2502}, \ref{eq25039}). 
The system of equations is efficiently solved by the Matlab ODE solver ODE23.

In combining the elements discussed above into a relevant model, a number of parameters 
are to be considered and set: 

$L$: The luminosity of the illuminating bright star (or sum of luminosities for an ensemble of such stars). 
Values of $L$ ranging from $10^{37}$ erg/s to $10^{40}$ erg/s have been
chosen. A flux of $10^{37}$ erg/s corresponds to a main-sequence star of 6-9 M$_\odot$, $10^{38}$ erg/s to 14-16 M$_\odot$, and $10^{39}$ to 35-80 M$_\odot$,
according to the mass-luminosity relations of \citet{Salaris05}, \citet{Eker15}, see also the evolution for a 60 M$_\odot$ model by \citet{Groh14}. If one adopts a
Salpeter IMF (\citet{Salpeter55}) with $N(m)\propto m^{-2.35}$, one finds that the summed-up luminosity distribution from a young stellar generation peaks
at stellar masses of $\sim20$ M$_\odot$, and that the light from stars around that value dominates over that from the much fewer stars with $m>$ 40 M$_\odot$. It also follows that the integrated luminosity from 30,000 stars (i.e. of the magnitude, including binaries, suggested by  \citet{Hurley05} to constitute the starting 
population of the cluster M67), distributed according to the Salpeter IMF produces a luminosity of $1.8\times 10^{39}$ erg/s.
The strong luminosity increase of the stars in connection with their supernova phases turns out to be of less importance, due to the short 
duration of those as compared with the main-sequence phases (we neglect the effects of the SN winds which are not thought to be very selective in their
effects on grains relative to gas). The luminosity of an individual star as well as its effective temperature may however vary 
significantly during the accreting Pre-Main-Sequence phase with time-scales as short as 100-1000 years (cf. \citet{Haemmerle16}), 
leading to variations in the number of photo-ionizing photons by about a factor of two which 
affects our $f$ parameter.

$f$: The fraction of the stellar flux that is emitted at wavelengths shorter than 91.2 nm, i.e. that is able to photo-ionize hydrogen atoms from the ground state. Values chosen here are 0.1 and 0.2, as estimated from the model atmospheres of hot stars by \citet{Castelli04}. We note,
however that according to the tables supplied by \citet{Rahner17}, values as high as 0.4 may be valid for early stellar populations. 

$t_{br}$: The characteristic time for the H\,II region around the still forming bright star to grow to the initial 
Str{\"o}mgren radius. As a standard, we 
set the parameter to 0, but also explored the effects of setting it to values up to 100,000 years. This, following \citet{Haemmerle16}, will probably overestimate the effects of the
delayed growth of the region.

$t_{end}$: The ending time for the evolution of the model system, here set to 6 Myr. During this time the central star/s, after an initial brightening,
is assumed to have a constant total luminosity. A value of 6 Myr is significantly longer than the life-times of the most massive stars, but since the stars with masses
around 20 solar masses collectively contribute most ionizing radiation, and the formation time of such stars in a cluster may be somewhat different, we have
chosen this value of $t_{end}$. Also, note that the cleansed mass $m_c(t)$ tends to level out beyond 1 Myr, see Figure \ref{fig4a}, which makes the choice of
$t_{end}$ less significant.       

$n_0$: The initial density in the gas determines the 
expansion speeds of the H\,II region and the dust front, as well as the rate of cleansing, $m_{c}(t)$. The density of H$_2$ molecules in the cloud, or alternatively of HI if a temperature of 100 K is chosen (see below), is usually estimated to be in the range 10$^2$-10$^5$ cm$^{-3}$ (\citet{Bergin07}),
a range reflecting the highly clumpy character of the medium. Different values in this range have been chosen alternatively.

$T_n$: The temperature of the neutral gas, alternatively chosen to be 10\,K and 100\,K, for an adopted composition of 
H$_2$ molecules and H atoms, respectively. $m_{\rm H}$ is changed from the mass of one hydrogen molecule to the mass of one hydrogen atom, accordingly.
As a standard, we have chosen the molecular alternative, i.e. $T_0=10$\,K and $m_{\rm H}=2\times 1.67\cdot 10^{-24}$\,g, and after that studied the consequences
of the alternative choice.  

$T_i$: The temperature of the gas with fully ionized hydrogen, is set to 10,000\,K.  

$\phi$: The gas/dust ratio by mass in star-forming regions is usually set to 100 (see, e.g. \citet{Liseau15}) and this value has been chosen here. 

$\mu$: The mass density of the gas, as normalized on the mass density of hydrogen, $\mu=1+(Y+Z)/X$, is set equal to 1.35.

$\rho_d$: The mass density of the dust grains is set to 1.0 g\,cm$^{-3}$.

\begin{figure}
  \resizebox{\hsize}{!}{\includegraphics{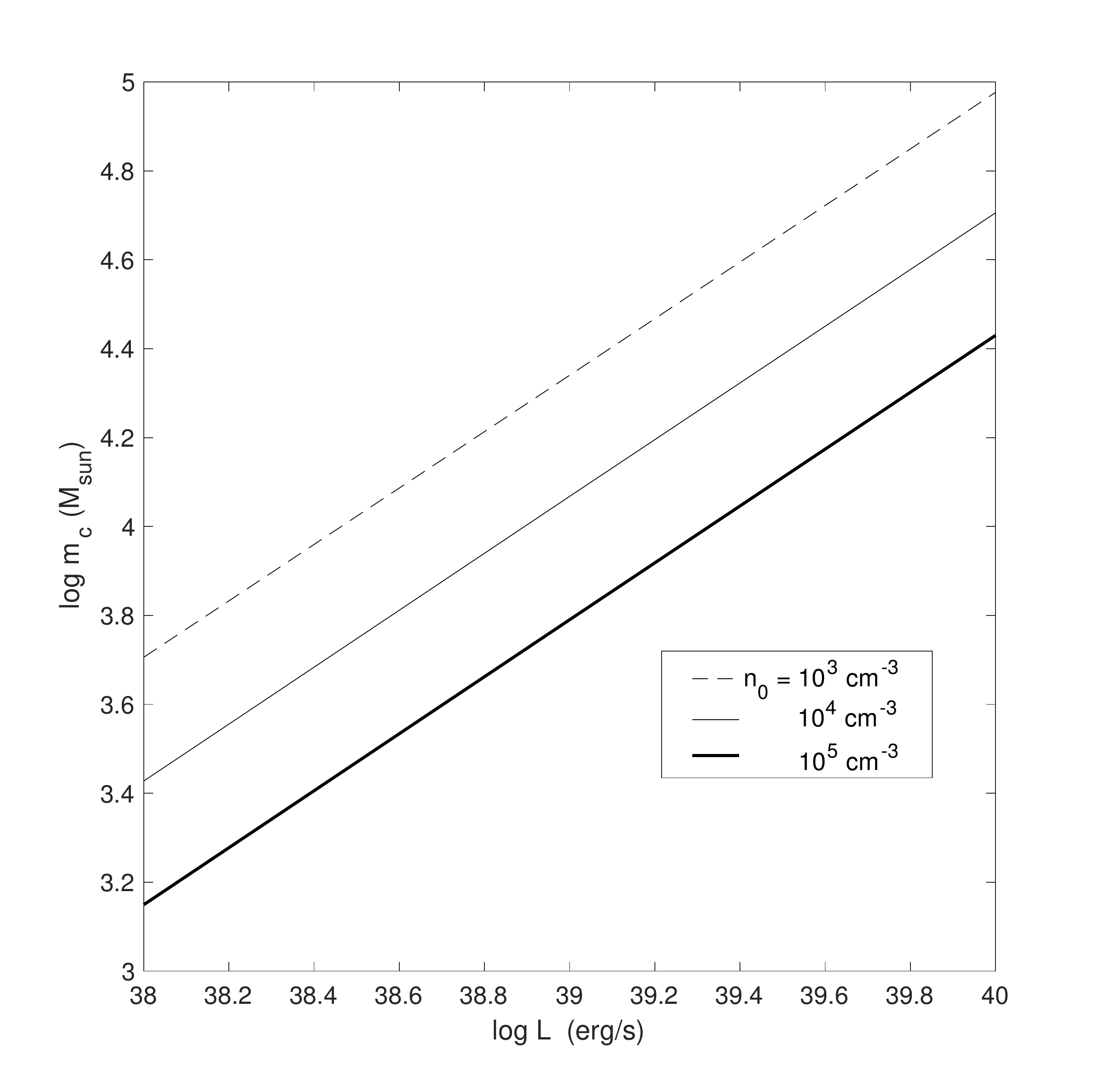}}
  \caption{The logarithmic cleansed mass, $m_{cl}(t)$, after 6 Gyr as a function of the luminosity of the central light source, for 
three different initial molecular number densities $n_0$.}
  \label{fig3a}
\end{figure}

\begin{figure}
  \resizebox{\hsize}{!}{\includegraphics{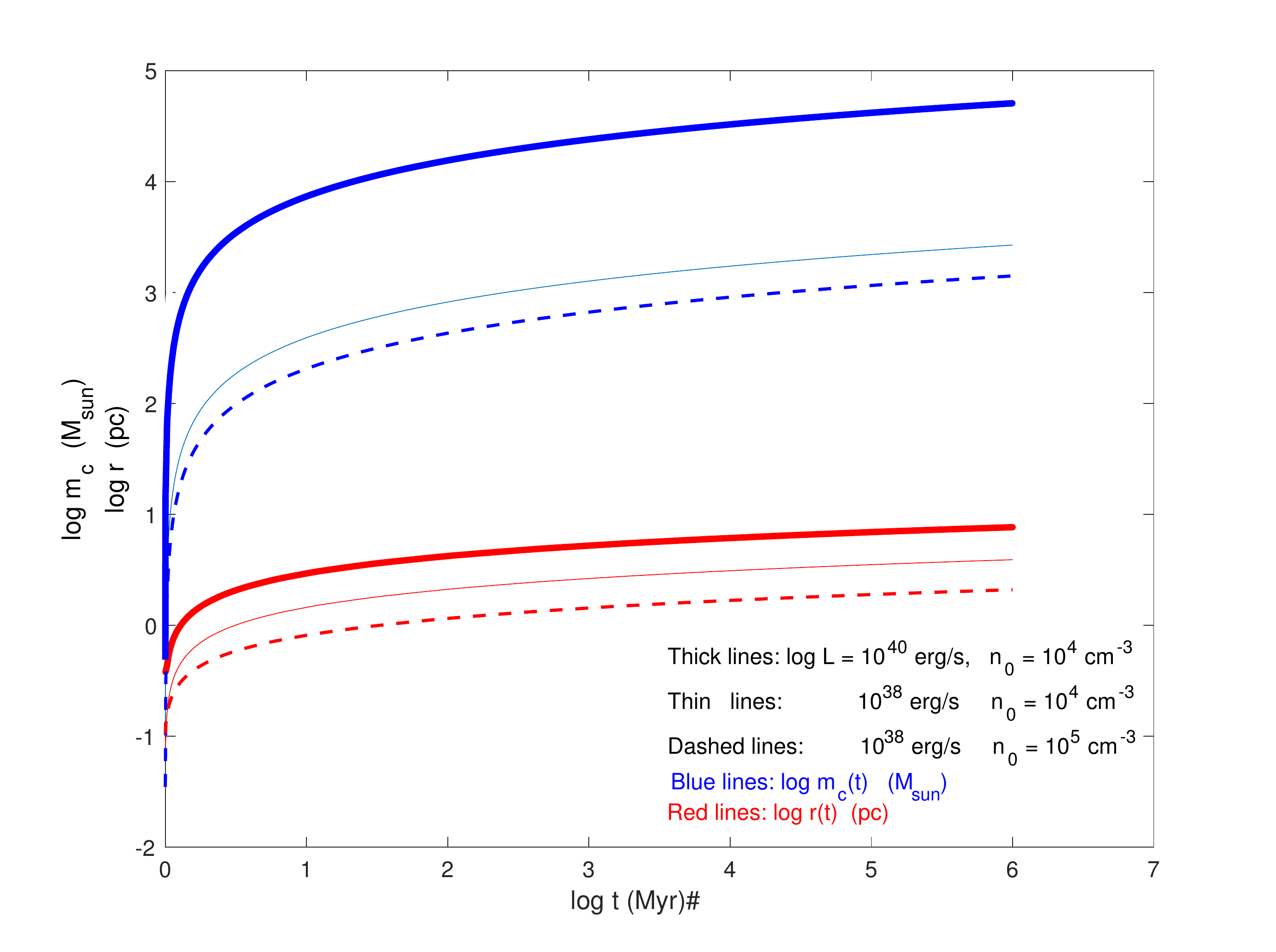}}
  \caption{The logarithmic cleansed mass, $m_{c}(t)$ (black and blue lines), and the radius of the dust front, $r(t)$ (red lines), as functions of time for 
two different luminosities of the central source, $L$, and two different initial molecular number densities $n_0$.}
  \label{fig4a}
\end{figure}

The maximum cleansed masses in 6 Myr, as a function of the stellar luminosity, are illustrated for different $n_0$ in Fig. \ref{fig3a}.
The time developments of the cleansed mass and the radius of the dust front are shown for some different choices of the parameters
in Fig. \ref{fig4a}. 

\section{Results and discussion}

\subsection{Results}

In Figure \ref{fig4a} is seen how the rapid growth of $m_{c}$ while the optical depth is small, decelerates 
when the dust front gets optically thick and then follows Equation (\ref{eq2506}). We can easily understand the main behavior in this
figure by approximating $r(t)$ by $R_{\rm H\,II}$ according to the Hosokawa \& Inutsuka approximation (see \citet{Bisbas15}, their
Equation (13)) which {\bf for $R_{St} \ll \mathcal{R}_{\rm I} \ll (T_i/T_0)^{3/4}$ leads to}
\begin{equation}
R_{\rm HII} \sim R_{St}^{3/7} t^{4/7}.
\label{eq88}
\end{equation}
Then, we find from the Equation (\ref{eq11}) and Equation (\ref{eq2506}) by integration
\begin{equation}
m_{cl}(t) \sim L^{1/2} n_0^{-1/4} \cdot t ,
\label{eq89}
\end{equation}
and
\begin{equation}
r(t) \sim n_0^{-2/7}\cdot t^{4/7} ,
\label{eq89}
\end{equation}
which agrees well with the tendencies in Figures \ref{fig3a} and \ref{fig4a}. It is interesting to note that the cleansed mass 
$m_{c}$ is only substantially affected by the density in the cloud via the effects of the initial density $n_0$ onto the radius of
the ionized region, and that an increasing density thus leads to a reduction of $m_{c}$.

\begin{figure}
  \resizebox{\hsize}{!}{\includegraphics{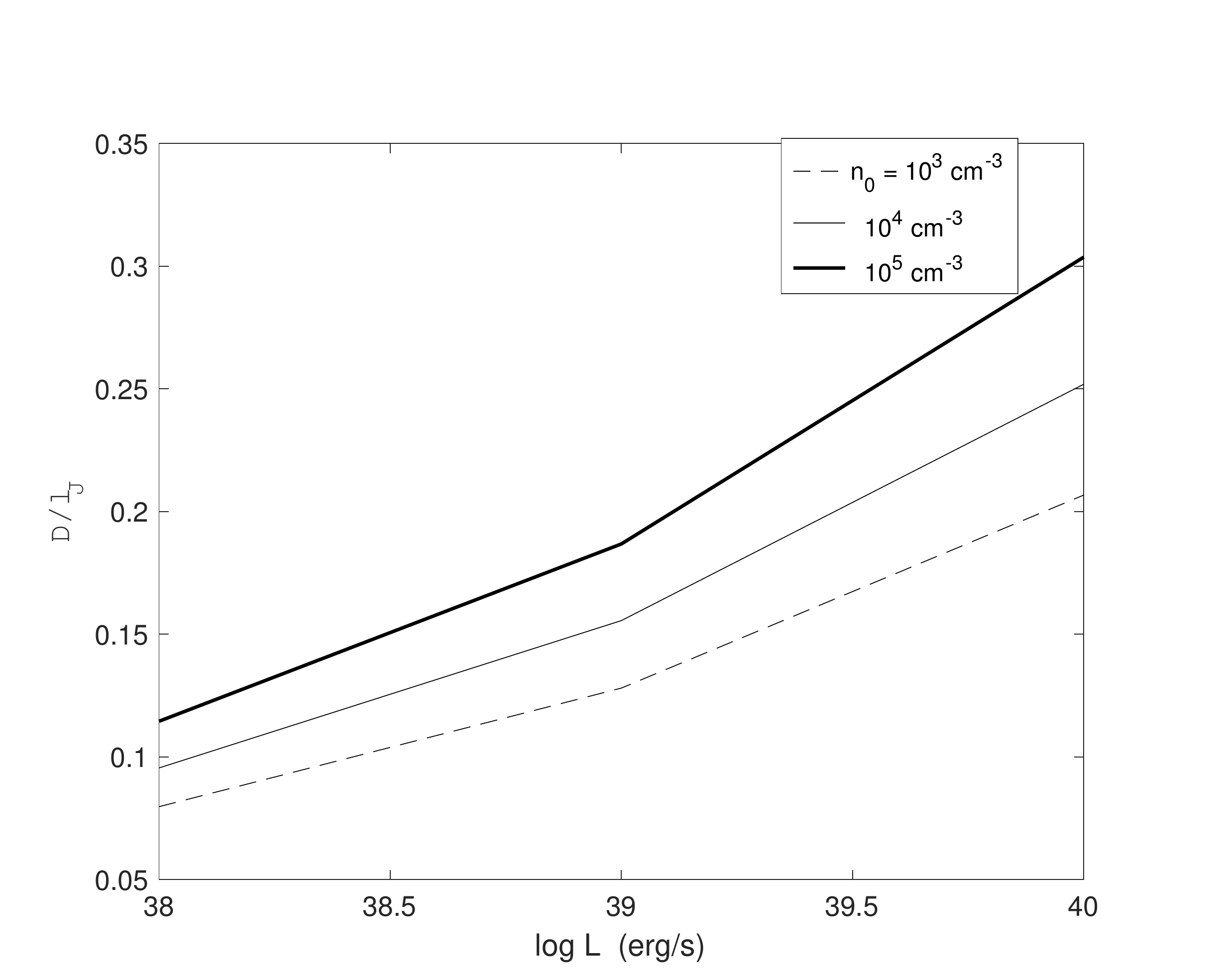}}
  \caption{The ratio of the thickness of the cleansed shell relative to the Jeans Length, $D/l_J$, for models with different central luminosities
$L$ and initial densities $n_0$.}
  \label{fig5a}
\end{figure}

A measure, of some but limited value of the ability for stars to form by gravitational instability 
out of the cleansed gas would be the thickness of 
the cleansed shell $D$ relative to the Jeans Length, $l_{\rm J}$,
\begin{equation}
l_{\rm J} = \sqrt{\frac{15 k T}{4\pi G \mu \rho}}.
\label{eq90}
\end{equation}
We have used the last term of Equation (\ref{eq25039}) to obtain $D$ and estimated the gas density $\rho(t)$ as
\begin{equation}
\rho(t) = \frac{m_{c}(t)}{4\pi(R_{\rm II} + D)^2 \cdot D}.
\label{eq91}
\end{equation} 
The results are illustrated in Fig. \ref{fig5a}. It is seen that the thickness of the
shell of cleansed gas is relatively small, as compared with the Jeans length, which seems to indicate that effective star formation in
this shell may be problematic.   

\subsection{Discussion}
It is clear from the {\bf results} above that an efficient production of cleansed gas is favoured by a high stellar luminosity $L$ and a relatively
low initial gas density $n_0$.  
In order to obtain the observed abundance effect of \citet{Melendez09} complete dust-cleansing is not
necessary; a reduction limited to about 20$\%$ of the refractories relative to volatiles would suffice. 
Thus, a cleansed mass of just 0.2 M$_\odot$ would in principle be sufficient in order to
produce the effect observed for the Sun, relative to solar twins in our neighbourhood in the Galaxy. However, this relaxing circumstance is 
counterbalanced by the fact that the star-fomation efficiency would probably not be much higher than 20$\%$ although local efficiencies of $50\%$ or more
cannot be excluded (see, e.g. \citet{Kim16} and references given there). 

It is important to consider whether the rough approximations behind the present estimates may have led to severe underestimates of the cleansed 
masses. 
One such approximation may be the early evolution of the bright star with a successive brightening
instead of, as supposed here, a direct switch on to a given constant $L$. We have tested the effects of a gradual brightening
of the star, from 1/10\,$L$ to $L$ in $10^5$ yr, but found only small effect on the cleansed masses, less than a few percent. The effect of our schematic
assumption of all the neutral gas to be molecular and at a low temperature was studied by adopting the alternative choice of {\it atomic} gas at 100\,K.
It was found that the reduction of $m_{\rm H}$ by a factor of two only caused small changes in $m_c(t)$ while the increase of the temperature typically reduced $m_c(t)$
by a factor of 2.2. This may also be seen directly from Equation (\ref{eq2506}). The possibility that the gas at the 
dust front and beyond that is molecular, while the cleansed gas is hotter and atomic, was also considered, again with small consequences.
More important are the
effects of inhomogeneities on the expansion rates of the H\,II regions, found by \citet{Bisbas15} to 
slow the mean expansion by up to $40\%$. This would lead to a systematic increase of the possible cleansed masses  
by up to a factor of about 2. 

\begin{figure}
  \resizebox{\hsize}{!}{\includegraphics{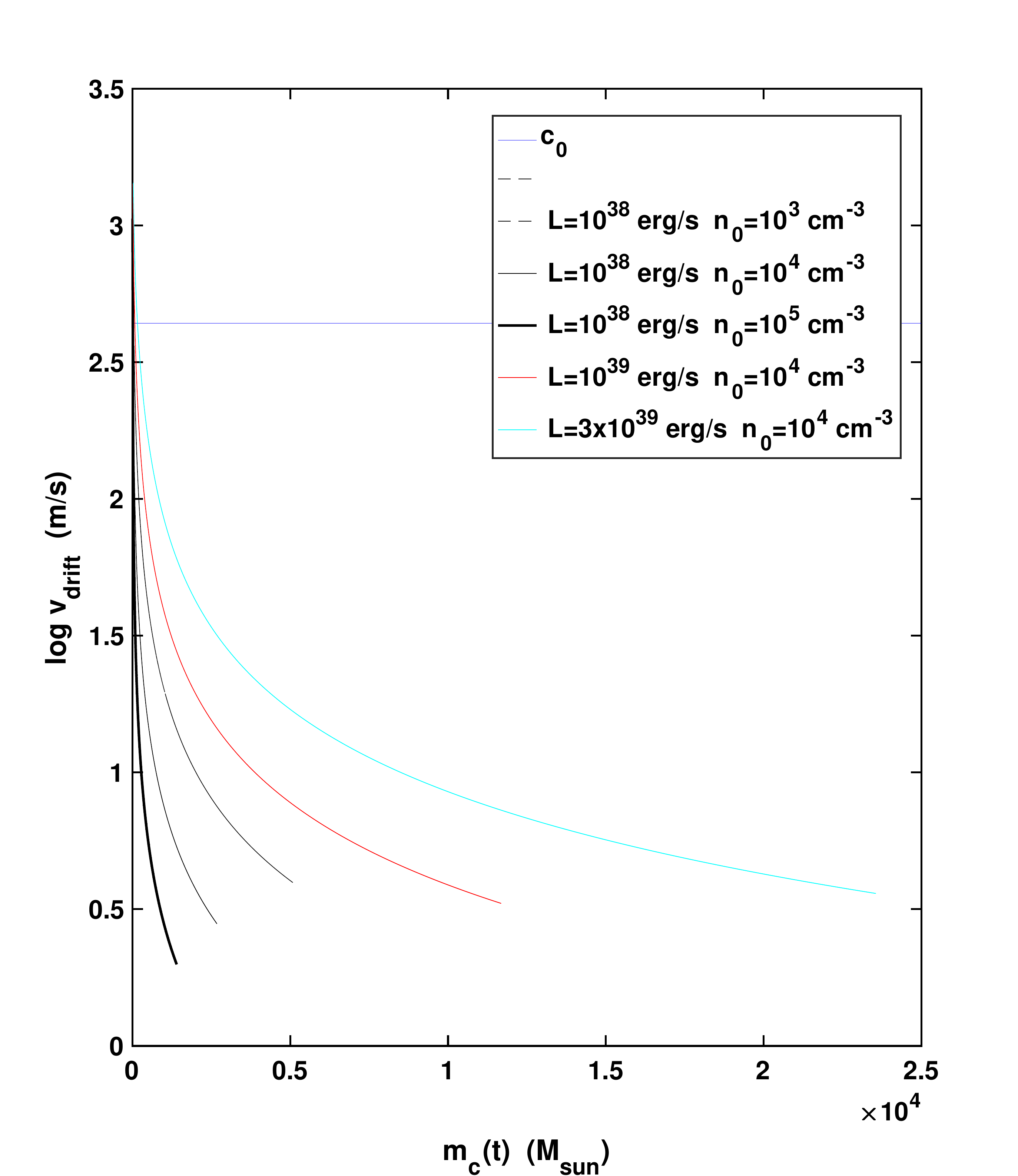}}
  \caption{The logarithmic drift velocity $v_{\rm drift}$ in m/s a function of the cleansed mass, $m_c(t)$ for a number of different stellar luminosities 
and initial number densities of H$_2$ molecules. The horizontal line indicates a typical value of the sound speed.}
  \label{fig51a}
\end{figure}

\begin{figure}
  \resizebox{\hsize}{!}{\includegraphics{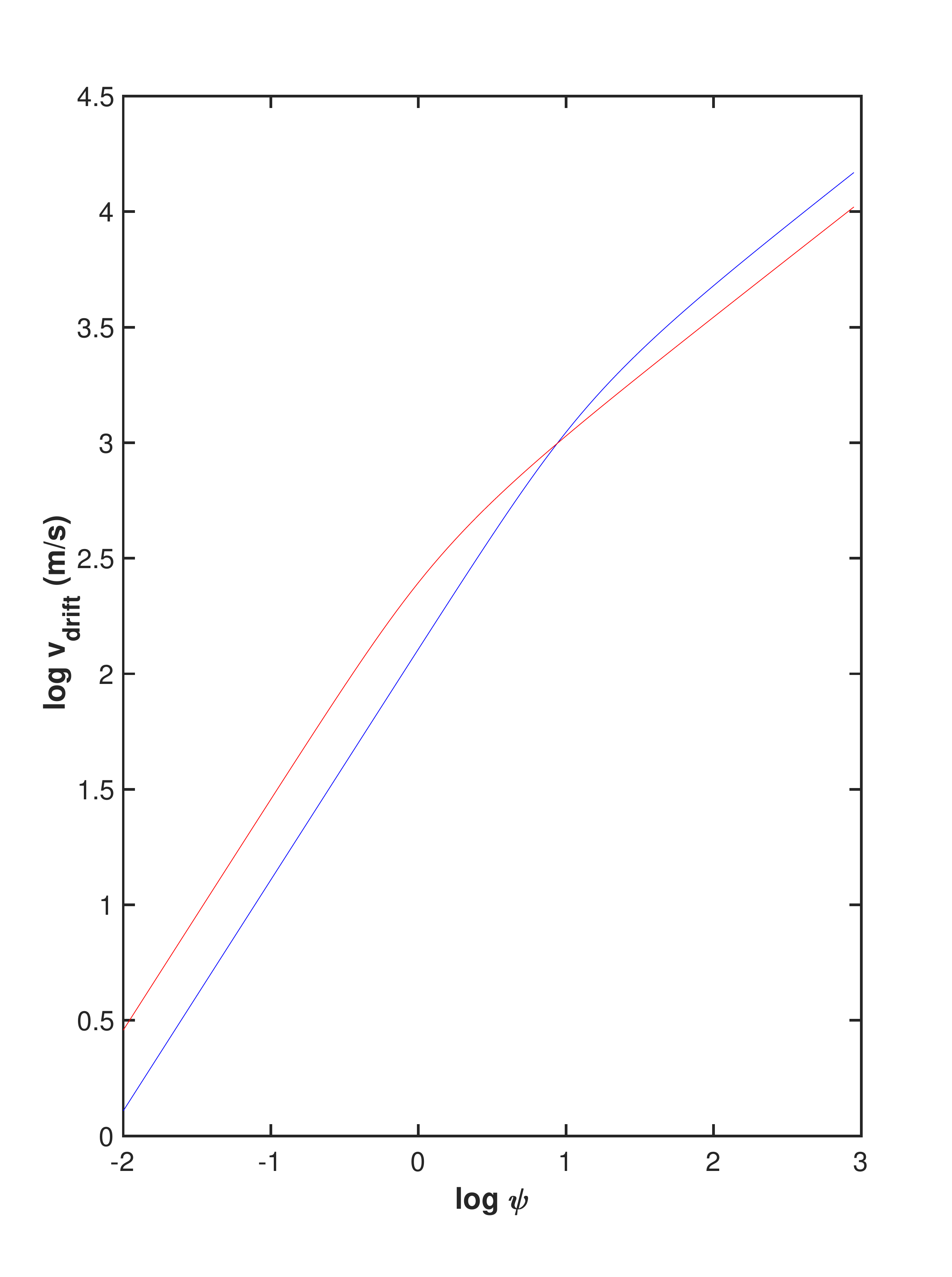}}
  \caption{The logarithmic drift velocity in m/s vs the quantity $\psi = L_{38}/d_{pc}^2 \cdot n_4^{-1}$ (for explanation, see text. Blue curve: atomic gas and $T_0 = 100$ K, red curve: molecular gas, $T_0 = 10$ K.}
  \label{fig52a}
\end{figure}

A severe problem for the cleansing mechanism studied here is that mixing 
may even out the differences gas/dust ratios between cleansed and un-cleansed regions. Turbulence with typical velocities of 1-2 km/s 
is observed in molecular clouds (see \citet{Hennebelle12} for a review). 
The extent to which turbulence is generated by self-gravity sooner than the result of e.g. radiative feed-back from
star-forming activity (for reviews, see \citet{Dobbs14} and \citet{Padoan14}, and also \citet{Raskutti16}) is still not clear. However, there are observational indications 
that turbulence may
be strong in regions of a cloud in early phases of star formation and even be the result of gravitational collapse (see \citet{Palau15}). It may also be driven by Rayleigh-Taylor and Kelvin-Helmholtz 
instabilities resulting from large-scale flows when the bubble of H\,II gas collides with the surrounding dense molecular gas, see \citet{Ochsendorf14}. 
Such turbulence could be supposed to
erase the effects of radiative cleansing and fractionation discussed here: the velocities caused by those latter effects are about one to two orders of magnitude
less than the observed turbulent velocities. As seen in Fig. \ref{fig51a} it is only in the very first phases, when the mass of the dust in the 
dust front is small and the front is close to the radiatively accelerating star/s, that the drift velocities are higher than 0.1 km/s, phases when $m_c(t)$ is
less than $10\%$ of its final value. It is clear from the Figure that the highest drift velocities, that could perhaps
compete with the turbulent mixing, would be in the light-luminosity low-density case, with $n_0\sim 10^2 \, {\rm cm}^{-3}$. If the dusty gas were still optically
thin {\bf one} might possibly expect high velocities even after a considerable gas mass were cleansed. Applying Equations (\ref{eq101}), (\ref{eq102}) and (\ref{eq201})
it is easy to see that the key factor is $L/(d^2 n)$
where $d$ is the distance from the star/s to the dust grain. Setting $\tau \approx 0$ one finds from Figure \ref{fig52a} that 
\begin{equation}
\psi = \frac{(L/10^{38}{\rm erg/s})}{(d/1{\rm{pc}})^2 (n/10^4 {\rm cm}^{-3})} > 10,
\label{eq2011}
\end{equation}
as the condition for $v_{\rm drift}$ to be equal or greater than 1 km/s in the optically thin case, for both the case of cool molecular and a somewhat hotter atomic hydrogen gas.
Obviously, this condition on $\psi$ may be fulfilled, e.g. by the early M\,67 cloud where $L$ may have been about $2\cdot 10^{39}$, and in particular
if $n$ would have been small. 

However, the requirement of a small optical depth is indeed not met for this scenario. In order to obtain a satisfactory yield of cleansed gas to explain the M\,67 abundances, one finds 
from Equation (\ref{eq1531}) that $\tau \sim 400 \, m_{c,4}/d_{pc}^2$, where
$m_{c,4}$ is the cleansed gas mass in units of $10^4$ solar masses. This estimate brings us back
to the estimates made above for optically thick dust fronts. Obviously, it is impossible to meet both the requirements of a high drift velocity on the
order of 1 km/s and a high cleansed mass.    
 
The question is then whether the small drift velocities, and the observed high turbulence velocities, 
make the radiative cleansing hypothesis untenable. 
If the minimum spatial scale of the turbulence is significantly smaller than our
typical values of the thickness of the cleansed gas shell $D$, on the order of 10,000 AU in typical cases,
the cleansing will be inhibited or at least reduced considerably. Although the spacial resolution in the observations are limited, there are
no signs in present data suggesting that this turbulence would be confined to the great scales. Instead, contemporary simulations suggest
smaller scales (\citet{Seifried17}).

It remains to explore whether efficient small-scale mixing close to the dust front, and further successive small-scale mixing in
the cleansed gas would be able to totally inhibit the effects of the cleansing. 

The Giant Molecular Clouds are known to have complex structures, with dense clumps of H$_2$ gas interfoliated
with regions with cool molecular gas with much lower densities (see further \citet{Hennebelle12}). 
If such a cloud is exposed to hot stellar radiation, the less dense regions may be
rapidly cleansed. 
One may argue that the inhomogeneities, not the least in the dust distribution (cf. e.g. \citet{Hopkins16}) could lead to instabilities in the neutral gas that could possibly amplify the radiative cleansing. 
 
Instabilities of various types are known
to occur during the evolution of H\,II regions, not the least at their outer boundaries (\citet{Garcia96}, \citet{Whalen08a}, \citet{Whalen08b}. Of a certain interest are here the ”resonant drag instabilities (RDI:s) of dust-gas mixtures 
(\citet{Squire17}), which are unstable at a wide range of spatial scales and rapidly growing, in particular when the
phase velocity of a wave in the gas agrees with the drift velocity component along that wave propagation vector (\citet{Hopkins17}). As pointed out by the referee of the present paper, these instabilities might affect the degrees of radiative cleansing. 
It should be noted, however, that the cleansed gas borders the dust front which
in itself is very over-rich in dust. It remains to study how great variations in gas/dust ratio that may occur in these 
regions, and in particular what the possibilities are for dust-cleansed gas to remain so.  

An argument against the effects of turbulent mixing of cleansed gas with more dust-rich one could be the existence of magnetic fields in the
gas. If the
fields were structured perpendicularly relative to the direction towards the illuminating stars, they could hamper turbulence provided that the ionization of the gas is non-negligible. However, such fields would also reduce the drift of the photoelectrically charged dust grains 
and thus diminish radiative cleansing considerably, see \citet{Draine11}. 
As recently shown by {\citet{Hopkins18}, the wealth of 
the newly discovered ”resonant drag instabilities” in magnetized dusty gases may, however, lead to modifications of this conclusion due
to more or less vastly growing magneto-sonic RDI:s}.
At any rate, the inhomogeneities and local instabilities discussed here 
would probably not contribute a systematic global dust cleansing of as much mass as needed for a cluster like M\,67.   

An interesting issue is whether the small possibilities for global radiative cleansing could be increased by a second gravitational collapse of the 
star-forming cloud, like that which may have affected the core of 30 Doradus where two different star generations have been found (\citet{Sabbi12}). 
\citet{Rahner17} have studied the circumstances under which re-collapse may occur and found that the star-formation efficiency must be below 
a few percent for low-mass clouds and below about 25 percent for clouds with masses on the order of $10^6$ M$_{\odot}$ to prevent the stellar feed-back due to the first generation of stars
from totally dispersing the cloud. The characteristic time from a star-forming episode to the next would be on the order of some free-fall times (i.e. 
3 - 6 Myr). If one could assume that the large-scale spherical symmetry could survive these episodes, so that the dust shell is not fully mixed into the 
collapsing cleansed gas within it, one might envisage a gradual dust-cleansing in several steps. For this to work, however, the effects of turbulence must
again be limited. It is highly questionable whether this is possible. Certainly, at a collapse the drift speed increases with the decreasing distance
$d$ to the cloud centre in proportion to $d^{-2}$, but it also decreases with an increasing gas density $n\sim d^3$ which at a uniform
contraction combines with the $d^{-2}$ dependence 
to a linear dependence of the drift speed on $d$ (c.f. Equation (\ref{eq1})). Thus, the drift speed in a cloud contracting towards
newly formed stars will in fact decrease to values far below turbulence velocities on the order of 1 km/s. 
    
\section{Conclusions}

Using simple arguments, essentially based on momentum conservation in a dusty {\bf neutral and} homogeneous gas affected by radiative pressure from nearby hot
stars, we have demonstrated that radiative cleansing of star-forming clouds might in principle explain composition differences between some single stars, such as the 
Sun as compared with solar twins. However, as is shown in Fig. \ref{fig5a} the cleansed gas shell is so geometrically thin that considerable inhomogeneities
are needed to induce gravitational instabilities of cleansed gas. In order to produce such abundance effects for a great fraction of stars in a rich cluster, such s those observed for M67 (\citet{Onehag11}, \citet{Onehag14}) an unrealistic radiative flux is required. As was discussed above, the luminosity of all the cluster stars in its youth is expected to be about $2\cdot10^{39}$ erg/s. 
This may produce, following the results presented in Figure \ref{fig4a}, at the most well cleansed gas for a few thousand stars. To this one should also apply the 
modifications due to the limits of $20\%$ cleansing observed for the refractory elements and the counteracting star forming efficiency, discussed above. Another difficulty is posed by the
relative chemical homogeneity found for the 14 cluster members measured by \citet{Onehag14}.  It is hard to understand why the progression of the dust
front in a molecular cloud, surrounding a massive association of hot and bright stars, would leave behind a gas, systematically cleansed from dust and corresponding heavy 
elements to the observed degree.  
One order of magnitude stronger cleansing than found in the present study is needed, or even more than that, since our estimates are believed to give upper boundaries to the real effects. We must conclude that the mechanism
discussed here does not explain the composition of M67. Also, 
shocks with density fluctuations, magnetic fields, and not the least turbulence should reduce the cleansing very considerably: the drift speeds of the dust relative
to the gas in most of our models are order of magnitude smaller than the observed turbulence velocities, which casts considerable doubt also on whether it is the real explanation for the odd solar composition. 

More detailed 3D hydrodynamical simulations of dust cleansing at the rim of expanding H\,II regions with due consideration of 
shocks and instabilities would be interesting. We do not, however, consider it probable that such simulations will lead to results which enhance the cleansing effects enough
to explain our observations of M67. The present 
result urges continued detailed studies of accurate stellar abundances in M67 and other rich clusters to verify our present observational finding of global 
enhancement of volatile elements (or missing refractories) in the cluster relative to the majority of field stars. 

Looking back at the development of this line of research, our finding in itself may be reflected on with some irony. The discovery of the solar-composition
peculiarity by \citet{Melendez09} stimulated various proposed explanations. One of these was the dust-cleansing hypothesis, which, in fact, was the
reason for our decision to study M67 in detail. We then thought that a demonstration of solar-like abundance anomalies for the cluster would be a clear
indication that the solar composition is the result of dust-cleansing. The anomaly was indeed found for the cluster by \citet{Onehag11} and \citet{Onehag14}
and later verified by \citet{Liu16}. 
But this fact, we now find, does not verify the hypothesis but just questions it: it turns out to be too difficult to cleanse enough of gas to get a sufficiently cleansed 
proto-cluster of this magnitude. So, we cannot explain the composition of M67, 
as the result of self-cleansing of the gas by the first hot stars in the forming cluster. Other possibilities, like those of the chemo-dynamical evolution of the 
Galaxy, have to be considered further. 

Since the discovery by \citet{Melendez09}, a trend contrary to the more or less uniform tendency for dust depletion of the M\,67 stars has been discovered: that 
several physical binary
stars show significant abundance differences between their ingoing components, with differences for the various elements correlated with the condensation temperature of the element in several cases. Notable such cases are the binaries 16 Cyg \citep{Tucci14, Nissen17}, XO-2N/XO-2S \citep{Ramirez15, Biazzo15}, WASP 94 A/B \citep{Teske16a}, HD\,133131 A/B \citep{Teske16b}, HAT-P-4 \citep{Saffe17}, and the co-moving pair HD240430/HD1240429 \citep{Oh17}. These findings certainly suggest more local effects than those explored here for M\,67. They are not in total disagreement with the hypothesis of dust-cleansing by bright stars; sooner the results of
the present study indicate that if such cleansing occurs, it would be more local as a result of the thinness of the cleansed shell, probably deformed by instabilities. 
Turbulent dust/gas separation, like that discussed by \citet{Liu16}, may possibly also play a role here. Some further efforts should thus be spent on
modelling the possible cleansing of individual smaller gas clouds, like the proto-solar cloud, by neighbouring hot stars. 
Similar simulations, in 3D with proper treatment of instabilities in the neighbourhood of the HII front and its 
preceding shock front expanding into a clumpy molecular cloud, are already carried out by several groups but might also include the radiative fields and turbulence
and their effects on the dust drift, in order to illuminate the cleansing possibilities. 

Other suggested reasons for the odd solar atmospheric composition, such as those presented in Sec. 1 including the early suggested planetary connection, as well as
the possibility of radiative self-cleansing from dust by protostars of their late accretion flows, 
should be further scrutinized.

\begin{acknowledgements}
Bengt Edvardsson, Kjell Eriksson, Bernd Freytag, Anders Johansen, Lars Mattsson, Jorge Mel{\'e}ndez, Dhruba Mitra, Poul Erik Nissen, and
an unknown referee, are thanked for many valuable suggestions and comments on the manuscript. 
\end{acknowledgements}

\bibliographystyle{aa} 
\bibliography{Gustafsson2018.bib}

\begin{thebibliography}{73}
\expandafter\ifx\csname natexlab\endcsname\relax\def\natexlab#1{#1}\fi

\bibitem[{{Adams}(2010)}]{Adams10}
{Adams}, F.~C. 2010, \araa, 48, 47

\bibitem[{{Adibekyan} {et~al.}(2017){Adibekyan}, {Delgado-Mena}, {Feltzing},
  {Gonz{\'a}lez Hern{\'a}ndez}, {Hinkel}, {Korn}, {Asplund}, {Beck}, {Deal},
  {Gustafsson}, {Honda}, {Lind}, {Nissen}, \& {Spina}}]{Adibekyan17}
{Adibekyan}, V., {Delgado-Mena}, E., {Feltzing}, S., {et~al.} 2017, ArXiv
  e-prints [\eprint[arXiv]{1701.05737}]

\bibitem[{{Adibekyan} {et~al.}(2014){Adibekyan}, {Gonz{\'a}lez Hern{\'a}ndez},
  {Delgado Mena}, {Sousa}, {Santos}, {Israelian}, {Figueira}, \& {Bertran de
  Lis}}]{Adibekyan14}
{Adibekyan}, V.~Z., {Gonz{\'a}lez Hern{\'a}ndez}, J.~I., {Delgado Mena}, E.,
  {et~al.} 2014, \aap, 564, L15

\bibitem[{{Akimkin} {et~al.}(2015){Akimkin}, {Kirsanova}, {Pavlyuchenkov}, \&
  {Wiebe}}]{Akimkin15}
{Akimkin}, V.~V., {Kirsanova}, M.~S., {Pavlyuchenkov}, Y.~N., \& {Wiebe}, D.~S.
  2015, \mnras, 449, 440

\bibitem[{{Akimkin} {et~al.}(2017){Akimkin}, {Kirsanova}, {Pavlyuchenkov}, \&
  {Wiebe}}]{Akimkin17}
{Akimkin}, V.~V., {Kirsanova}, M.~S., {Pavlyuchenkov}, Y.~N., \& {Wiebe}, D.~S.
  2017, \mnras, 469, 630

\bibitem[{{Arthur} {et~al.}(2004){Arthur}, {Kurtz}, {Franco}, \&
  {Albarr{\'a}n}}]{Arthur04}
{Arthur}, S.~J., {Kurtz}, S.~E., {Franco}, J., \& {Albarr{\'a}n}, M.~Y. 2004,
  \apj, 608, 282

\bibitem[{{Baraffe} \& {Chabrier}(2010)}]{Baraffe10}
{Baraffe}, I. \& {Chabrier}, G. 2010, \aap, 521, A44

\bibitem[{{Bergin} \& {Tafalla}(2007)}]{Bergin07}
{Bergin}, E.~A. \& {Tafalla}, M. 2007, \araa, 45, 339

\bibitem[{{Biazzo} {et~al.}(2015){Biazzo}, {Gratton}, {Desidera}, {Lucatello},
  {Sozzetti}, {Bonomo}, {Damasso}, {Gandolfi}, {Affer}, {Boccato}, {Borsa},
  {Claudi}, {Cosentino}, {Covino}, {Knapic}, {Lanza}, {Maldonado}, {Marzari},
  {Micela}, {Molaro}, {Pagano}, {Pedani}, {Pillitteri}, {Piotto}, {Poretti},
  {Rainer}, {Santos}, {Scandariato}, \& {Zanmar Sanchez}}]{Biazzo15}
{Biazzo}, K., {Gratton}, R., {Desidera}, S., {et~al.} 2015, \aap, 583, A135

\bibitem[{{Bisbas} {et~al.}(2015){Bisbas}, {Haworth}, {Williams}, {Mackey},
  {Tremblin}, {Raga}, {Arthur}, {Baczynski}, {Dale}, {Frostholm}, {Geen},
  {Haugb{\o}lle}, {Hubber}, {Iliev}, {Kuiper}, {Rosdahl}, {Sullivan}, {Walch},
  \& {W{\"u}nsch}}]{Bisbas15}
{Bisbas}, T.~G., {Haworth}, T.~J., {Williams}, R.~J.~R., {et~al.} 2015, \mnras,
  453, 1324

\bibitem[{{Castelli} \& {Kurucz}(2004)}]{Castelli04}
{Castelli}, F. \& {Kurucz}, R.~L. 2004, ArXiv Astrophysics e-prints
  [\eprint{astro-ph/0405087}]

\bibitem[{{Casuso} \& {Beckman}(2010)}]{Casuso10}
{Casuso}, E. \& {Beckman}, J.~E. 2010, \aj, 139, 1406

\bibitem[{{Cochran} \& {Ostriker}(1977)}]{Cochran77}
{Cochran}, W.~D. \& {Ostriker}, J.~P. 1977, \apj, 211, 392

\bibitem[{{Diaz-Miller} {et~al.}(1998){Diaz-Miller}, {Franco}, \&
  {Shore}}]{Diaz-Miller98}
{Diaz-Miller}, R.~I., {Franco}, J., \& {Shore}, S.~N. 1998, \apj, 501, 192

\bibitem[{{Dobbs} {et~al.}(2014){Dobbs}, {Krumholz}, {Ballesteros-Paredes},
  {Bolatto}, {Fukui}, {Heyer}, {Low}, {Ostriker}, \&
  {V{\'a}zquez-Semadeni}}]{Dobbs14}
{Dobbs}, C.~L., {Krumholz}, M.~R., {Ballesteros-Paredes}, J., {et~al.} 2014,
  Protostars and Planets VI, 3

\bibitem[{{Draine}(2011{\natexlab{a}})}]{Draine11}
{Draine}, B.~T. 2011{\natexlab{a}}, \apj, 732, 100

\bibitem[{{Draine}(2011{\natexlab{b}})}]{Draine11b}
{Draine}, B.~T. 2011{\natexlab{b}}, {Physics of the Interstellar and
  Intergalactic Medium}

\bibitem[{{Eker} {et~al.}(2015){Eker}, {Soydugan}, {Soydugan}, {Bilir}, {Yaz
  G{\"o}k{\c c}e}, {Steer}, {T{\"u}ys{\"u}z}, {{\c S}eny{\"u}z}, \&
  {Demircan}}]{Eker15}
{Eker}, Z., {Soydugan}, F., {Soydugan}, E., {et~al.} 2015, \aj, 149, 131

\bibitem[{{Franco} {et~al.}(1991){Franco}, {Ferrini}, {Barsella}, \&
  {Ferrara}}]{Franco91}
{Franco}, J., {Ferrini}, F., {Barsella}, B., \& {Ferrara}, A. 1991, \apj, 366,
  443

\bibitem[{{Franco} {et~al.}(1990){Franco}, {Tenorio-Tagle}, \&
  {Bodenheimer}}]{Franco90}
{Franco}, J., {Tenorio-Tagle}, G., \& {Bodenheimer}, P. 1990, \apj, 349, 126

\bibitem[{{Fujimoto} {et~al.}(2018){Fujimoto}, {Krumholz}, \&
  {Tachibana}}]{Fujimoto18}
{Fujimoto}, Y., {Krumholz}, M.~R., \& {Tachibana}, S. 2018, ArXiv e-prints
  [\eprint[arXiv]{1802.08695}]

\bibitem[{{Garcia-Segura} \& {Franco}(1996)}]{Garcia96}
{Garcia-Segura}, G. \& {Franco}, J. 1996, \apj, 469, 171

\bibitem[{{Gonzalez} {et~al.}(2010){Gonzalez}, {Carlson}, \&
  {Tobin}}]{Gonzalez10}
{Gonzalez}, G., {Carlson}, M.~K., \& {Tobin}, R.~W. 2010, \mnras, 407, 314

\bibitem[{{Groh} {et~al.}(2014){Groh}, {Meynet}, {Ekstr{\"o}m}, \&
  {Georgy}}]{Groh14}
{Groh}, J.~H., {Meynet}, G., {Ekstr{\"o}m}, S., \& {Georgy}, C. 2014, \aap,
  564, A30

\bibitem[{{Gustafsson} {et~al.}(2010){Gustafsson}, {Mel{\'e}ndez}, {Asplund},
  \& {Yong}}]{Gustafsson10}
{Gustafsson}, B., {Mel{\'e}ndez}, J., {Asplund}, M., \& {Yong}, D. 2010, \apss,
  328, 185

\bibitem[{{Haemmerl{\'e}} \& {Peters}(2016)}]{Haemmerle16}
{Haemmerl{\'e}}, L. \& {Peters}, T. 2016, \mnras, 458, 3299

\bibitem[{{Hennebelle} \& {Falgarone}(2012)}]{Hennebelle12}
{Hennebelle}, P. \& {Falgarone}, E. 2012, \aapr, 20, 55

\bibitem[{{Hopkins} \& {Lee}(2016)}]{Hopkins16}
{Hopkins}, P.~F. \& {Lee}, H. 2016, \mnras, 456, 4174

\bibitem[{{Hopkins} \& {Squire}(2017)}]{Hopkins17}
{Hopkins}, P.~F. \& {Squire}, J. 2017, ArXiv e-prints
  [\eprint[arXiv]{1707.02997}]

\bibitem[{{Hopkins} \& {Squire}(2018)}]{Hopkins18}
{Hopkins}, P.~F. \& {Squire}, J. 2018, ArXiv e-prints
  [\eprint[arXiv]{1801.10166}]

\bibitem[{{Hurley} {et~al.}(2005){Hurley}, {Pols}, {Aarseth}, \&
  {Tout}}]{Hurley05}
{Hurley}, J.~R., {Pols}, O.~R., {Aarseth}, S.~J., \& {Tout}, C.~A. 2005,
  \mnras, 363, 293

\bibitem[{{Kim} {et~al.}(2016){Kim}, {Kim}, \& {Ostriker}}]{Kim16}
{Kim}, J.-G., {Kim}, W.-T., \& {Ostriker}, E.~C. 2016, \apj, 819, 137

\bibitem[{{Krumholz} \& {Matzner}(2009)}]{Krumholz09}
{Krumholz}, M.~R. \& {Matzner}, C.~D. 2009, \apj, 703, 1352

\bibitem[{{Liseau} {et~al.}(2015){Liseau}, {Larsson}, {Lunttila}, {Olberg},
  {Rydbeck}, {Bergman}, {Justtanont}, {Olofsson}, \& {de Vries}}]{Liseau15}
{Liseau}, R., {Larsson}, B., {Lunttila}, T., {et~al.} 2015, \aap, 578, A131

\bibitem[{{Liu} {et~al.}(2016){Liu}, {Asplund}, {Yong}, {Mel{\'e}ndez},
  {Ram{\'{\i}}rez}, {Karakas}, {Carlos}, \& {Marino}}]{Liu16}
{Liu}, F., {Asplund}, M., {Yong}, D., {et~al.} 2016, \mnras, 463, 696

\bibitem[{{Lopez} {et~al.}(2014){Lopez}, {Krumholz}, {Bolatto}, {Prochaska},
  {Ramirez-Ruiz}, \& {Castro}}]{Lopez14}
{Lopez}, L.~A., {Krumholz}, M.~R., {Bolatto}, A.~D., {et~al.} 2014, \apj, 795,
  121

\bibitem[{{Maldonado} {et~al.}(2015){Maldonado}, {Eiroa}, {Villaver},
  {Montesinos}, \& {Mora}}]{Maldonado15}
{Maldonado}, J., {Eiroa}, C., {Villaver}, E., {Montesinos}, B., \& {Mora}, A.
  2015, \aap, 579, A20

\bibitem[{{Maldonado} \& {Villaver}(2016)}]{Maldonado16}
{Maldonado}, J. \& {Villaver}, E. 2016, \aap, 588, A98

\bibitem[{{Mart{\'{\i}}nez-Gonz{\'a}lez}
  {et~al.}(2014){Mart{\'{\i}}nez-Gonz{\'a}lez}, {Silich}, \&
  {Tenorio-Tagle}}]{Martinez-Gonzalez14}
{Mart{\'{\i}}nez-Gonz{\'a}lez}, S., {Silich}, S., \& {Tenorio-Tagle}, G. 2014,
  \apj, 785, 164

\bibitem[{{Mathews}(1967)}]{Mathews67}
{Mathews}, W.~G. 1967, \apj, 147, 965

\bibitem[{{Mathis} {et~al.}(1977){Mathis}, {Rumpl}, \& {Nordsieck}}]{Mathis77}
{Mathis}, J.~S., {Rumpl}, W., \& {Nordsieck}, K.~H. 1977, \apj, 217, 425

\bibitem[{{Mel{\'e}ndez} {et~al.}(2009){Mel{\'e}ndez}, {Asplund}, {Gustafsson},
  \& {Yong}}]{Melendez09}
{Mel{\'e}ndez}, J., {Asplund}, M., {Gustafsson}, B., \& {Yong}, D. 2009, \apjl,
  704, L66

\bibitem[{{Nissen}(2015)}]{Nissen15}
{Nissen}, P.~E. 2015, \aap, 579, A52

\bibitem[{{Nissen} {et~al.}(2017){Nissen}, {Silva Aguirre},
  {Christensen-Dalsgaard}, {Collet}, {Grundahl}, \& {Slumstrup}}]{Nissen17}
{Nissen}, P.~E., {Silva Aguirre}, V., {Christensen-Dalsgaard}, J., {et~al.}
  2017, ArXiv e-prints [\eprint[arXiv]{1710.03544}]

\bibitem[{{Ochsendorf} {et~al.}(2014){Ochsendorf}, {Verdolini}, {Cox},
  {Bern{\'e}}, {Kaper}, \& {Tielens}}]{Ochsendorf14}
{Ochsendorf}, B.~B., {Verdolini}, S., {Cox}, N.~L.~J., {et~al.} 2014, \aap,
  566, A75

\bibitem[{{Oh} {et~al.}(2017){Oh}, {Price-Whelan}, {Brewer}, {Hogg}, {Spergel},
  \& {Myles}}]{Oh17}
{Oh}, S., {Price-Whelan}, A.~M., {Brewer}, J.~M., {et~al.} 2017, ArXiv e-prints
  [\eprint[arXiv]{1709.05344}]

\bibitem[{{{\"O}nehag} {et~al.}(2014){{\"O}nehag}, {Gustafsson}, \&
  {Korn}}]{Onehag14}
{{\"O}nehag}, A., {Gustafsson}, B., \& {Korn}, A. 2014, \aap, 562, A102

\bibitem[{{{\"O}nehag} {et~al.}(2011){{\"O}nehag}, {Korn}, {Gustafsson},
  {Stempels}, \& {Vandenberg}}]{Onehag11}
{{\"O}nehag}, A., {Korn}, A., {Gustafsson}, B., {Stempels}, E., \&
  {Vandenberg}, D.~A. 2011, \aap, 528, A85

\bibitem[{{Padoan} {et~al.}(2014){Padoan}, {Federrath}, {Chabrier}, {Evans},
  {Johnstone}, {J{\o}rgensen}, {McKee}, \& {Nordlund}}]{Padoan14}
{Padoan}, P., {Federrath}, C., {Chabrier}, G., {et~al.} 2014, Protostars and
  Planets VI, 77

\bibitem[{{Palau} {et~al.}(2015){Palau}, {Ballesteros-Paredes},
  {V{\'a}zquez-Semadeni}, {S{\'a}nchez-Monge}, {Estalella}, {Fall}, {Zapata},
  {Camacho}, {G{\'o}mez}, {Naranjo-Romero}, {Busquet}, \& {Fontani}}]{Palau15}
{Palau}, A., {Ballesteros-Paredes}, J., {V{\'a}zquez-Semadeni}, E., {et~al.}
  2015, \mnras, 453, 3785

\bibitem[{{Pellegrini} {et~al.}(2007){Pellegrini}, {Baldwin}, {Brogan},
  {Hanson}, {Abel}, {Ferland}, {Nemala}, {Shaw}, \& {Troland}}]{Pellegrini07}
{Pellegrini}, E.~W., {Baldwin}, J.~A., {Brogan}, C.~L., {et~al.} 2007, \apj,
  658, 1119

\bibitem[{{Rahner} {et~al.}(2017){Rahner}, {Pellegrini}, {Glover}, \&
  {Klessen}}]{Rahner17}
{Rahner}, D., {Pellegrini}, E.~W., {Glover}, S.~C.~O., \& {Klessen}, R.~S.
  2017, \mnras, 470, 4453

\bibitem[{{Ram{\'{\i}}rez} {et~al.}(2015){Ram{\'{\i}}rez}, {Khanal}, {Aleo},
  {Sobotka}, {Liu}, {Casagrande}, {Mel{\'e}ndez}, {Yong}, {Lambert}, \&
  {Asplund}}]{Ramirez15}
{Ram{\'{\i}}rez}, I., {Khanal}, S., {Aleo}, P., {et~al.} 2015, \apj, 808, 13

\bibitem[{{Ram{\'{\i}}rez} {et~al.}(2009){Ram{\'{\i}}rez}, {Mel{\'e}ndez}, \&
  {Asplund}}]{Ramirez09}
{Ram{\'{\i}}rez}, I., {Mel{\'e}ndez}, J., \& {Asplund}, M. 2009, \aap, 508, L17

\bibitem[{{Raskutti} {et~al.}(2016){Raskutti}, {Ostriker}, \&
  {Skinner}}]{Raskutti16}
{Raskutti}, S., {Ostriker}, E.~C., \& {Skinner}, M.~A. 2016, \apj, 829, 130

\bibitem[{{Sabbi} {et~al.}(2012){Sabbi}, {Lennon}, {Gieles}, {de Mink},
  {Walborn}, {Anderson}, {Bellini}, {Panagia}, {van der Marel}, \& {Ma{\'{\i}}z
  Apell{\'a}niz}}]{Sabbi12}
{Sabbi}, E., {Lennon}, D.~J., {Gieles}, M., {et~al.} 2012, \apjl, 754, L37

\bibitem[{{Saffe} {et~al.}(2017){Saffe}, {Jofr{\'e}}, {Martioli}, {Flores},
  {Petrucci}, \& {Jaque Arancibia}}]{Saffe17}
{Saffe}, C., {Jofr{\'e}}, E., {Martioli}, E., {et~al.} 2017, \aap, 604, L4

\bibitem[{{Salaris} \& {Cassisi}(2005)}]{Salaris05}
{Salaris}, M. \& {Cassisi}, S. 2005, {Evolution of Stars and Stellar
  Populations}, 400

\bibitem[{{Salpeter}(1955)}]{Salpeter55}
{Salpeter}, E.~E. 1955, \apj, 121, 161

\bibitem[{{Schal{\'e}n}(1939)}]{Schalen39}
{Schal{\'e}n}, C. 1939, \zap, 17, 260

\bibitem[{{Seifried} {et~al.}(2017){Seifried}, {Walch}, {Girichidis}, {Naab},
  {W{\"u}nsch}, {Klessen}, {Glover}, {Peters}, \& {Clark}}]{Seifried17}
{Seifried}, D., {Walch}, S., {Girichidis}, P., {et~al.} 2017, \mnras, 472, 4797

\bibitem[{{Shu} {et~al.}(2002){Shu}, {Lizano}, {Galli}, {Cant{\'o}}, \&
  {Laughlin}}]{Shu02}
{Shu}, F.~H., {Lizano}, S., {Galli}, D., {Cant{\'o}}, J., \& {Laughlin}, G.
  2002, \apj, 580, 969

\bibitem[{{Skinner} \& {Ostriker}(2015)}]{Skinner15}
{Skinner}, M.~A. \& {Ostriker}, E.~C. 2015, \apj, 809, 187

\bibitem[{{Spina} {et~al.}(2016){Spina}, {Mel{\'e}ndez}, \&
  {Ram{\'{\i}}rez}}]{Spina16}
{Spina}, L., {Mel{\'e}ndez}, J., \& {Ram{\'{\i}}rez}, I. 2016, \aap, 585, A152

\bibitem[{{Spitzer}(1941)}]{Spitzer41}
{Spitzer}, Jr., L. 1941, \apj, 94, 232

\bibitem[{{Squire} \& {Hopkins}(2017)}]{Squire17}
{Squire}, J. \& {Hopkins}, P.~F. 2017, ArXiv e-prints
  [\eprint[arXiv]{1711.03975}]

\bibitem[{{Str{\"o}mgren}(1939)}]{Stromgren39}
{Str{\"o}mgren}, B. 1939, \apj, 89, 526

\bibitem[{{Teske} {et~al.}(2016{\natexlab{a}}){Teske}, {Khanal}, \&
  {Ram{\'{\i}}rez}}]{Teske16a}
{Teske}, J.~K., {Khanal}, S., \& {Ram{\'{\i}}rez}, I. 2016{\natexlab{a}}, \apj,
  819, 19

\bibitem[{{Teske} {et~al.}(2016{\natexlab{b}}){Teske}, {Shectman}, {Vogt},
  {D{\'{\i}}az}, {Butler}, {Crane}, {Thompson}, \& {Arriagada}}]{Teske16b}
{Teske}, J.~K., {Shectman}, S.~A., {Vogt}, S.~S., {et~al.} 2016{\natexlab{b}},
  \aj, 152, 167

\bibitem[{{Tucci Maia} {et~al.}(2014){Tucci Maia}, {Mel{\'e}ndez}, \&
  {Ram{\'{\i}}rez}}]{Tucci14}
{Tucci Maia}, M., {Mel{\'e}ndez}, J., \& {Ram{\'{\i}}rez}, I. 2014, \apjl, 790,
  L25

\bibitem[{{Walch} {et~al.}(2012){Walch}, {Whitworth}, {Bisbas}, {W{\"u}nsch},
  \& {Hubber}}]{Walch12}
{Walch}, S.~K., {Whitworth}, A.~P., {Bisbas}, T., {W{\"u}nsch}, R., \&
  {Hubber}, D. 2012, \mnras, 427, 625

\bibitem[{{Whalen} \& {Norman}(2008{\natexlab{a}})}]{Whalen08b}
{Whalen}, D. \& {Norman}, M.~L. 2008{\natexlab{a}}, \apj, 673, 664

\bibitem[{{Whalen} \& {Norman}(2008{\natexlab{b}})}]{Whalen08a}
{Whalen}, D.~J. \& {Norman}, M.~L. 2008{\natexlab{b}}, \apj, 672, 287

\end{thebibliography}
\end{document}